\documentclass[12pt,openany]{article}

\usepackage{amsfonts,latexsym,eucal,color,graphicx,epsfig,amssymb,amsmath}

\begin{document}
\begin{titlepage}

\vspace{2cm}
\begin{center}
{\Large {\bf Measurement errors in entanglement-assisted electron microscopy}} \\
\vspace{2cm}
Hiroshi Okamoto\\
\vspace{.5cm} 
{\it Department of Electronics and Information Systems, 

Akita PrefecturalUniversity, Yurihonjo 015-0055, Akita, Japan}\\
\vspace{.5cm}
Electronic address: okamoto@akita-pu.ac.jp\\
\vspace{1.5cm}
\vspace{.5cm}
\vspace{1.5cm}
\end{center}

\begin{abstract}
The major resolution-limiting factor in cryoelectron microscopy of
unstained biological specimens is radiation damage by the very electrons
that are used to probe the specimen structure. To address this problem,
an electron microscopy scheme that employs quantum entanglement to
enable phase measurement precision beyond the standard quantum limit
has recently been proposed {[}Phys. Rev. A \textbf{85}, 043810{]}.
Here we identify and examine in detail measurement errors that will
arise in the scheme. An emphasis is given to considerations concerning
inelastic scattering events because in general schemes assisted with
quantum entanglement are known to be highly vulnerable to lossy processes.
We find that the amount of error due both to elastic and inelastic
scattering processes are acceptable provided that the electron beam
geometry is properly designed.
\end{abstract}
\end{titlepage}
\renewcommand{\thesection}{\Roman{section}}
\renewcommand{\thesubsection}{\Alph{subsection}}
\renewcommand{\thesubsubsection}{\arabic{subsubsection}}

\section{Introduction}

In cryoelectron microscopy, unstained biological specimens are rapidly
vitrified at cryogenic temperatures \cite{Glaeser Taylor 1978,Adrian Dubochet Nature}.
Consequently, artefacts due to heavy-metal staining, desiccation,
and other sample preparation processes are no longer an issue. However,
the frozen, hydrated biological specimen, consisting mostly of light
elements, scatters electrons weakly. Hence biological specimens generally
are weak phase objects associated with low image contrast \cite{Comment 1}.
In this setting, the resolution is limited by radiation damage by
the probe electrons \cite{Henderson Review} to approximately 5-10
nm in the case of \emph{single} objects \cite{CryoET prospect}. This
leaves much to be desired because 2 nm resolution would be needed
to identify molecules in frozen vitrified slices of the cell in cryoelectron
tomography \cite{CryoET review}, or 0.8 nm resolution would be required
to observe the secondary structure of a single protein molecule. The
reason why radiation damage limits the resolution is that the 'safe'
electron dose, which does not cause sizable damage to the specimen,
is so small that the low-contrast image is dominated by shot noise.
Shot noise is a manifestation of the particle nature of the electron
and hence is fundamental.

Several approaches to address the radiation damage problem are known.
First, methods based on averaging, such as two-dimensional crystallography
\cite{e-crystallography book} and single-particle analysis \cite{Frank single particle},
represent an established branch of methodology in structural biology.
In favorable cases, these methods essentially attained atomic resolution
\cite{Yu Jin Zhou}. However, in order to average out the noise, this
approach requires at least thousands of copies of the molecule of
interest without much structural variance and hence is not suited
for soft or unique objects. Second, the advent of in-focus phase contrast
electron microscopy \cite{Danev Nagayama,Majorovits et al,Cambie phase contrast}
enabled researchers to see weak phase objects much clearer than hitherto
possible. The reason is that it provides a well-behaving phase contrast
transfer function (CTF) that does not fall to zero at low resolutions
and does not oscillate at high resolutions. However, this method does
not go beyond the standard quantum limit, as will be mentioned. Third
and finally, the use of low acceleration voltage down to, e. g. 20
kV reduces radiation damage. In particular, one may choose the electron
energy below the onset of relevant knock-on damage thresholds, although
the dominant damage mechanism in biological cryoelectron microscopy
is considered to be ionization events due to radiolysis \cite{radiation damage mechanism}.
There have recently been much effort on developing aberration-corrected
low-voltage transmission electron microscopes (TEMs) and scanning
TEMs (STEMs) around the globe \cite{JEOL LV AC TEM,NION LV AC TEM,SALVE LV AC TEM}.
While this approach certainly makes sense, the inelastic mean free
path is considerably shorter at lower electron energies, necessitating
preparation of ultra-thin specimens \cite{Comment 2}. While there
have been reports of impressive images of \emph{thin} biological molecules
\cite{Fujiyoshi DNA} and organic molecules \cite{Koshino-Suenaga}
since decades ago, hydrated macromolecules of biological interest
can have the size more than 50 nm. This makes it necessary to study
thicker specimens and hence, if we insist on studying biologically
interesting frozen hydrated macromolecules or vitrified slices of
the cell, the use of a sufficiently high acceleration voltage is necessary.
Overall, these considerations suggest that no current electron microscopy
method produces, in a robust and widely applicable manner, images
of single hydrated objects of biological interest at a resolution
below 2 nm. Furthermore, phase measurement in all electron microscopy
methods developed thus far, including all mentioned above, is governed
by the shot noise limit, or in other words, the standard quantum limit.
In this case, the precision $\delta\varphi$ of the measurement of
the small phase shift $\varphi$, associated with the biological specimen,
scales with the number of electrons $N$ as $\delta\varphi\sim1/\sqrt{N}$. 

That the standard quantum limit can be beaten is well known in the
field of quantum metrology \cite{qMetrology review}. While the field
has begun decades ago \cite{Helstrom book,squeezed}, a sizable fraction
of recent activities in quantum metrology, mostly in the context of
optics, revolves around the idea of employing entangled quantum states
\cite{Yurke,ent-assisted atomic clock,ent-assisted lithography}.
In particular, measurement of a phase shift $\delta\varphi$ is a
familiar objective in quantum metrology and it is relevant also to
biological electron microscopy where weak phase objects are dealt
with. A major question in quantum metrology is how the measurement
precision $\delta\varphi$ varies with the amount of relevant 'resources'
$N$. Among others, 'query complexity' emerged as one of the most
useful 'resource count' \cite{query complexity}, which has been further
elucidated and shown to be governed by the Heisenberg limit $\delta\varphi\sim1/N$
\cite{limit qMetrology}. Roughly, the number of queries equals the
number of interactions with the entity to be measured. Hence query
complexity nicely captures the resource in biological electron microscopy,
where the experimenter wants to get the most out of each 'query',
which corresponds to passing of each electron in the specimen. It
is worth noting here that, notwithstanding the recent theoretical
\cite{beating Heisenberg theory} and experimental \cite{beating Heisenberg Experiment}
reports of beating the Heisenberg limit, as long as the number of
query is taken as the resource count, the Heisenberg limit does represent
the fundamental limit \cite{limit qMetrology}.

It should be noted that the Heisenberg limit is by no means an easy
target in real situations \cite{Elusive Heisenberg}. It has increasingly
been recognized that entanglement-assisted measurement is vulnerable
to lossy processes and a number of recent studies address this problem
\cite{lossy qMet-1,lossy qMet-2,lossy qMet-3,lossy qMet-4}. It is
now established that quantum metrology, in lossy situations, offers
'merely' a constant-factor improvement in the measurement precision,
as opposed to the quadratic asymptotic improvement by a factor proportional
to $\sqrt{N}$. Nevertheless, a constant-factor improvement is all
that we want in biological electron microscopy, or perhaps in any
measurement for that matter. The real question is whether the value
of the constant factor enables relevant specific resolutions, such
as aforementioned 2 nm or 0.8 nm. This requires an analysis specific
to biological cryoelectron microscopy, but in doing so, we may learn
lessons of more general character.

On a related note, a recent work \cite{multipixel qMet} proposes
an efficient multi-pixel phase estimation method that takes advantage
of quantum entanglement. While interesting, the scheme employs a delocalized
probe state over all the pixels and hence it is unlikely to be robust
against localizing lossy processes, which occur in biological electron
microscopy. 

It should also be noted that, in general, beating the standard quantum
limit does not necessarily require an entangled quantum state \cite{Luis}.
Repeated use of a probe particle would suffice. However, in the present
case of entanglement-assisted biological electron microscopy, entangled
quantum states are necessary for a somewhat mundane reason that the
repeated use of an electron would result in too large a scattering
angle to be handled by practical electron optics.

The use of quantum advantage specifically in the context of biological
electron microscopy has been discussed for some time now, albeit mostly
from the theoretical perspective \cite{Okamoto-1,Okamoto-2,Glaeser physics today,Putnam-Yanik,Okamoto-3,ent-assisted EM}.
In this paper we build on the recently proposed scheme that uses quantum
entanglement between the probe electron and the Cooper pair box (CPB)
placed on an electron mirror \cite{ent-assisted EM}. The scheme realizes
what effectively amounts to multiparticle entanglement between the
probe electrons, by way of sequential interactions between each electron
and the CPB. A major objective of this paper is to investigate how
and to what extent the effect of lossy processes may be mitigated
in the proposed scheme.

A distinct quantum approach \cite{Putnam-Yanik} to biological electron
microscopy based on interaction-free measurement \cite{Elitzur-Vaidman,Kwiat et al},
also suggested in Ref. \cite{Glaeser physics today}, has been proposed.
While interesting, we note that there is certain limit associated
to this type of approach \cite{Greame Mitchison}.

This paper is organized as follows. We first review the entanglement-assisted
electron microscopy scheme in Sec. \ref{sec:Review-ent-EM}. This
is followed by Sec. \ref{sec:Coherent errors}, in which we deal with
errors that is not related to lossy processes. In Sec. \ref{sec:Effect-of-inelastic},
we study how inelastic scattering processes make adverse effects to
the scheme. We also show that an appropriate design of the electron
probe geometry enables us to combat this type of effect. Sec. \ref{sec:Conclusion}
concludes the paper. Symbols $e,m$ respectively denotes the absolute
value of the electron charge and the electron mass. A \emph{diffraction
plane} refers to any plane conjugate to the back focal plane of the
objective lens. Since aberration-corrected electron optics is becoming
a commonplace, we generally neglect lens aberrations unless explicitly
stated otherwise. The probe electron energy is assumed to be $300\mathrm{keV}$
throughout the paper.

\section{Electron microscopy assisted by quantum entanglement\label{sec:Review-ent-EM}}

\subsection{A brief review\label{sub:A-brief-review}}

\begin{figure}
\begin{center}
\includegraphics[width=0.45\textwidth]{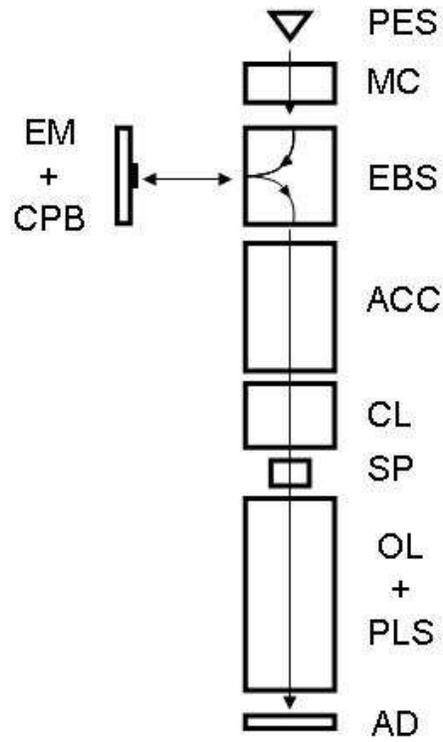}\caption{\label{fig:ent-asst-TEM structure}Structure of the proposed transmission
electron microscope. Electrons emitted form a pulsed electron source
(PES) goes through a monochromator (MC) and are directed by an electron
beam separator (EBS) towards a low temperature electron mirror (EM),
on which a Cooper pair box (CPB) is placed. All these processes happen
with a very low kinetic energy of the electron. Afterwards, the electron
is accelerated in an accelerator (ACC), goes through a condenser lens
(CL), specimen (SP), objective lens (OL), projector lens system (PLS)
and finally detected by an area detector (AD). The OL and PLS may
not be necessary in actual settings.}

\end{center}
\end{figure}

Here we briefly review the microscope proposed in Ref. \cite{ent-assisted EM},
in part because we also want to fix notations \cite{Comment 3}. The
scheme takes advantage of superconducting quantum electronics that
includes a single CPB \cite{Cooper pair box}, presumably in the circuit
quantum electrodynamics configuration \cite{circuitQED}. The microscope
is designed to measure the difference $\Delta\varphi=\varphi_{1}-\varphi_{0}$
of phase shifts $\varphi_{0}$ and $\varphi_{1}$ that are respectively
associated with two regions $S_{0}$ and $S_{1}$ on the biological
specimen, which generally is a weak phase object. Multitude of such
a measurement form an image. On the face of it, the shapes of the
two regions $S_{0}$ and $S_{1}$ may be chosen arbitrarily: For example,
they may be adjacent two small pixels, or it may be that $S_{0}$
surrounds a smaller region $S_{1}$. However, we will find in the
present work that the latter choice is much better. The CPB is placed
on the surface of an electron mirror, which in turn is inserted between
the pulsed electron gun and the condenser lens so that the electron
emitted from the electron gun first interacts with the CPB before
going to the specimen (See Fig. \ref{fig:ent-asst-TEM structure}).
We use only two states $|0>_{b}$ and $|1>_{b}$ of the CPB, that
respectively denote quantum states of the CPB with zero or one excess
Cooper pair, where the subscript 'b' stands for the word 'box'. The
main function of the electron mirror in this scheme is to modify the
electron beam shape, depending on the charging state of the CPB: The
electron mirror is designed in such a way that the beam trajectory
is so sensitive to the CPB charge state that only a single Cooper
pair in the CPB is sufficient to direct the electron beam to the specimen
region $S_{1}$ that would otherwise go to $S_{0}$. The scattered
electrons then go through more or less ordinary electron optics and
are detected at a plane, which is not conjugate to the image plane,
by an area detector with high quantum efficiency. Alternatively, unlike
the case of Ref. \cite{ent-assisted EM}, we will assume that the
probe electrons are detected on a diffraction plane in the present
work. In this case, it may well be possible to get rid of the objective
lens (OL) and the projector lens system (PLS) shown in Fig. \ref{fig:ent-asst-TEM structure}
altogether, resulting in an instrument resembling the scanning TEM
(STEM).

We mention several engineering challenges involved. A more complete
discussion of most of what follows has been presented elsewhere \cite{ent-assisted EM}.
There are mainly five aspects that need to be developed. First, while
the electron mirror has long history, an electron mirror that works
at temperatures generated by the dilution refrigerator, i.e. a few
tens of millikelvin, must be developed. Second, a pulsed electron
gun \cite{pulsed electron gun}, with $\sim\mathrm{meV}$ energy spread
and ps pulse width, must be developed. Each pulse may or may not contain
an electron for the scheme to work. We note that $\sim\mathrm{meV}$
energy spread has already been realized in the field of high resolution
electron energy loss spectroscopy (HREELS) where very low energy ($\sim1\mathrm{eV}$)
electrons are dealt with \cite{HREELS}. In a similar fashion, the
part of the electron optical system shown in Fig. \ref{fig:ent-asst-TEM structure},
consisting of the pulsed electron source, monochromator, electron
beam separator and the electron mirror, controls electrons with such
low kinetic energy. Unlike the case of TEM-EELS, we do \emph{not}
need a several hundred keV high voltage source with the voltage stability
comparable to the energy spread ($\sim1\mathrm{meV}$) of the electron
beam. Third, the Cooper pair box on the electron mirror must be operated
in a controlled fashion and such operations must be synchronized with
the pulsed electron gun. Fourth, as noted earlier, the electron optical
system must be sensitive enough so that a single excess Cooper pair
in the Cooper pair box can be 'seen' with the electron optics. Fifth
and finally, after interacting with the Cooper pair box, the probe
electron must be accelerated to several hundred keV before interacting
with the biological specimen. In principle, this can be done by brute
force, i.e. by electrically floating a large part of the electron
microscope on either side of the accelerator ACC shown in Fig. \ref{fig:ent-asst-TEM structure}
by several hundred kilovolts \cite{ent-assisted EM}. A recent development
in this connection, on the other hand, is the use of radio-frequency
accelerator that has been applied to TEMs \cite{Yang RF-TEM,Nagatani RF-TEM}
and it seems to have good compatibility with a pulsed electron beam.
If this approach moves beyond the proof-of-principle stage, then it
will be unnecessary to electrically float such a large part of the
microscope as described above.

We describe the \emph{single electron process} that lies at the heart
of our method. As shown later, mutiple single electron processes ($k$
of them) will comprise a \emph{k-electron process} in our scheme.
The single electron process involves a single imaging electron and
proceeds as follows. Suppose that the CPB is in a superposed charge
state \begin{equation}
|\psi_{1}>_{b}=c_{0}|0>_{b}+c_{1}|1>_{b}\label{eq:pre CPB state}\end{equation}
before the electron from the electron gun approaches the CPB. After
the electron interacts with the CPB, it gets entangled with the CPB
so that the whole quantum state of the CPB-electron system is\begin{equation}
|\Psi_{1}>=c_{0}|0>_{b}\otimes|0>+c_{1}|1>_{b}\otimes|1>\label{eq:pre whole state}\end{equation}
where the electron states $|0>$ and $|1>$ respectively denotes electron
waves incident on $S_{0}$ and $S_{1}$ regions on the specimen. Upon
transmission through the specimen, the electron wave in the state
$|1>$ acquires a phase factor $e^{i\Delta\varphi}$ relative to the
state $|0>$, corresponding to the phase shift $\Delta\varphi$. Hence
the state becomes $|\Psi_{2}>=c_{0}|0>_{b}\otimes|0>+c_{1}e^{i\Delta\varphi}|1>_{b}\otimes|1>$.
Let the electron state $|d_{j}>$ represent a diffracted wave to the
$j$-th pixel of the area detector. We expand the electron states
$|0>,|1>$ in terms of these states as\[
|0>=\sum_{j}a_{j}|d_{j}>,\textrm{ }|1>=\sum_{j}b_{j}|d_{j}>.\]
Using these, we obtain another expression of the state $|\Psi_{2}>$
as\[
|\Psi_{2}>=\sum_{j}\left(c_{0}a_{j}|0>_{b}+c_{1}b_{j}e^{i\Delta\varphi}|1>_{b}\right)\otimes|d_{j}>.\]
Upon detection of the electron at the $j$-th pixel of the area detector,
the CPB is left in the state \begin{equation}
|\psi_{2}>_{b}=\frac{1}{F}\left(c_{0}a_{j}|0>_{b}+c_{1}b_{j}e^{i\Delta\varphi}|1>_{b}\right),\label{eq:post CPB state-1}\end{equation}
where $F$ is a real normalization factor. 

In reality, an electron wave is not a qubit. One way to make the above
argument somewhat more precise is to expand the state $|1>$ as$ $\[
|1>=|1a>+|1b>+|1c>+\cdots\]
where each state on the right hand side is spatially more localized
on the specimen than the state $|1>$ is. The state becomes, after
transmission through the specimen,\[
e^{i\Delta\varphi_{a}}|1a>+e^{i\Delta\varphi_{b}}|1b>+e^{i\Delta\varphi_{c}}|1c>+\cdots\]
In this case, the above phase shift value $\Delta\varphi$ would represent
something similar to, but not identical with, the average of $\Delta\varphi_{a},\Delta\varphi_{b},\cdots$.
This way of representation would be more suitable when we deal with
a specimen with a structure within the region of the electron beam.
However, we will not pursue this representation. The point is also
related to the \emph{similar intensity map condition} \cite{ent-assisted EM},
to which we now turn.

To simplify the analysis, we assume that $\left|a_{j}\right|\cong\left|b_{j}\right|$,
which amounts to saying that the electron waves $|0>,|1>$ from the
two small regions $S_{0},S_{1}$ have similar intensity on the area
detector. This assumption is what we call the similar intensity map
condition. The assumption appears to be natural, but to what extent
the similar intensity map condition is satisfied must be carefully
evaluated, which will be done in Sec. \ref{sec:Coherent errors}.
Once we accept $\left|a_{j}\right|\cong\left|b_{j}\right|$, we may
write $b_{j}=a_{j}e^{i\beta_{j}}$, where $\beta_{j}$ represents
a \emph{known} phase shift associated with the optical path length
difference between two electron trajectories, starting respectively
at the specimen regions $S_{0},S_{1}$ and ending at the $j$-th pixel
of the area detector. The CPB state (\ref{eq:post CPB state-1}) may
then be expressed as, up to the overall phase factor,\begin{equation}
|\psi_{2}>_{b}=\frac{\left|a_{j}\right|}{F}\left(c_{0}|0>_{b}+c_{1}e^{i\left(\Delta\varphi+\beta_{j}\right)}|1>_{b}\right)=c_{0}|0>_{b}+c_{1}e^{i\left(\Delta\varphi+\beta_{j}\right)}|1>_{b}.\label{eq:post CPB state-2}\end{equation}
Since we know $\beta_{j}$, by quantum information processing machinery
developed for the CPB \cite{circuitQED} we may nullify it to obtain\begin{equation}
|\psi_{3}>_{b}=c_{0}|0>_{b}+c_{1}e^{i\Delta\varphi}|1>_{b}.\label{eq:post CPB state-3}\end{equation}
Let us call this last step as the \emph{phase compensation step}.
Comparing equations (\ref{eq:pre CPB state}) and (\ref{eq:post CPB state-3}),
we see that, assuming that the similar intensity map condition is
satisfied, the net effect induced by the single electron process is
multiplication of the phase factor $e^{i\Delta\varphi}$ to the CPB
state $|1>_{b}$ relative to $|0>_{b}$.

The k-electron process in our scheme consists of, in the order of
execution, the CPB initialization step, repeated single electron processes
that is described above ($k$ times), and the CPB readout step. First,
the CPB is initialized to the state $|\psi_{0}>_{b}=\left(|0>_{b}+|1>_{b}\right)/\sqrt{2}$.
Second, the single electron process is repeated $k$ times, resulting
in the state\begin{equation}
|\psi_{4}>_{b}=\frac{|0>_{b}+e^{ik\Delta\varphi}|1>_{b}}{\sqrt{2}}.\label{eq:CPB state after repeated unit processes}\end{equation}
This state may be expressed as, up to the overall phase factor, \[
|\psi_{4}>_{b}=\frac{-i}{\sqrt{2}}\left(\cos\frac{k\Delta\varphi}{2}+\sin\frac{k\Delta\varphi}{2}\right)|\leftarrow>_{b}+\frac{1}{\sqrt{2}}\left(\cos\frac{k\Delta\varphi}{2}-\sin\frac{k\Delta\varphi}{2}\right)|\rightarrow>_{b},\]
where $|\leftarrow>_{b}\equiv\left(|0>_{b}+i|1>_{b}\right)/\sqrt{2}$
and $|\rightarrow>_{b}\equiv\left(|0>_{b}-i|1>_{b}\right)/\sqrt{2}$.
Finally, in the CPB readout step we measure the CPB state with respect
to the basis $|\leftarrow>_{b},|\rightarrow>_{b}$. Probabilities
to find the state in these basis states are respectively\begin{equation}
P_{\leftarrow}=\frac{1+\sin\left(k\Delta\varphi\right)}{2},\textrm{ }P_{\rightarrow}=1-P_{\leftarrow}.\label{eq:CPB reading prob}\end{equation}
Notice, by the way, that the $\beta_{j}$ correction step between
the states (\ref{eq:post CPB state-2}) and (\ref{eq:post CPB state-3})
within the k-electron process may collectively be postponed until
right before the CPB readout step. In this case, all $k$ phase correction
values are recorded; and their sum is used in the combined correction
step.

We show how the Heisenberg limit can be approached. We write the electron
dose for a measurement on a single spot on the specimen $N$, which
must be kept below a certain value in order not to damage the specimen
significantly. Hence we can repeat the k-electron process $n=N/k$
times (that we assume to be an integer for simplicity). As is well
known from the theory of binomial distribution, the random variable
$X$ representing the number of finding the CPB state in the state
$|\leftarrow>_{b}$ has the expectation value $E\left[X\right]=nP_{\leftarrow}$
and variance $Var\left[X\right]=nP_{\leftarrow}P_{\rightarrow}$.
To simplify the analysis, we assume that $k\Delta\varphi\ll1$, although
this may not necessarily be a good approximation in some of actual
settings. Then we obtain $P_{\leftarrow}\cong\left(1+k\Delta\varphi\right)/2$
and $Var\left[X\right]=n/4+O\left(nk^{2}\Delta\varphi^{2}\right)\cong n/4$.
Introducing another random variable $Y=\left(2X/n-1\right)/k$, we
have\[
E\left[Y\right]=\Delta\varphi,\textrm{ }Var\left[Y\right]=\frac{4}{k^{2}n^{2}}Var\left[X\right]=\frac{1}{kN}.\]
Hence, the expected error in determining $\Delta\varphi$ with the
k-electron process is $\approx1/\sqrt{kN}$. Note that this represents
the standard quantum limit when $k=1$ and the Heisenberg limit when
$k=N$.

\subsection{Several remarks\label{sub:Several-remarks}}

We remark that, despite being sequential, our scheme may also be regarded
as one using something akin to the bosonic NOON state \cite{NOON state}.
Suppose that we had an unusually long electron microscope column between
the CPB-incorporated electron mirror and the specimen. Then all $k$
electrons would be flying simultaneously in the column before any
of them colliding with the specimen. From the discussions above, we
see that these electrons would be in the state\[
\frac{1}{\sqrt{2}}\left(|0>_{b}\otimes|00\cdots0>+|1>_{b}\otimes|11\cdots1>\right),\]
where $|00\cdots>$ denotes the state in which all electrons are in
the state $|0>$, and likewise for $|11\cdots1>$.

There are at least three sources of errors in the scheme reviewed
in this section. First, to what extent the similar intensity map condition
holds should be investigated, as remarked above. If this condition
is significantly violated, then the absolute values of the coefficients
$c_{0},c_{1}$ in equation (\ref{eq:pre CPB state}) deviate from
the ideal $1/\sqrt{2}$, making the 'contrast' of the final measurement
weaker. Second, when inelastic scattering happens, which is indeed
$2$ times more likely to happen than the elastic counterpart in cryoelectron
microscopy \cite{e-crystallography book,Wall Isaacson & Langmore},
we lose some coherence in the CPB state. In particular, if the inelastic
event is \emph{localized}, then the specimen would 'measure' the probe
electron state with respect to the localized basis states $|0>,|1>$.
In this case, the coherence would be completely lost. However, the
actual situation is subtler and we will investigate how things should
behave. Third, we have not thoroughly analyzed the CPB-electron entangling
interaction to estimate what and how much errors would be introduced
when the electron bounces off the CPB in the electron mirror. We will
consider the first two sources of errors in this paper, while leaving
the last item for future analysis.

What frequency of faulty measurement can we tolerate? To get a rough
idea, let us simplify the situation and consider a totally destructive
event that projects the CPB state onto either $|0>_{b}$ or $|1>_{b}$.
Suppose that such a destructive event occurs at random with the probability
$p_{d}$. The number of k-electron process is $n=N/k$ and any k-electron
process fails if a destructive event occurs more than once during
$k$ single electron processes. (For simplicity, we assume that the
experimenter, being somewhat lazy, does not abort the k-electron process
upon such a destructive event.) The success probability of a k-electron
process is $\left(1-p_{d}\right)^{k}\cong e^{-kp_{d}}$ if $p_{d}$
is sufficiently small. The number of k-electron processes for the
same dose $N$ is $n'=ne^{-kp_{d}}$. Retracing the argument in the
previous subsection \ref{sub:A-brief-review}, we have the expectation
value $E\left[X'\right]=n'P_{\leftarrow}$, variance $Var\left[X'\right]=n'P_{\leftarrow}P_{\rightarrow}$,
$Y'=\left(2X'/n'-1\right)/k$, and eventually\begin{equation}
E\left[Y'\right]=\Delta\theta,\textrm{ }Var\left[Y'\right]=\frac{1}{k^{2}n'}=\frac{1}{kNe^{-kp_{d}}}.\end{equation}
We want the variance $Var\left[Y'\right]$ smaller than the variance
in the conventional case $\sim N^{-1}$. To find $k=k_{m}$ that minimizes
the ratio $Var\left[Y'\right]/N^{-1}$, we differentiate this expression
and equate the result to zero. We find $k_{m}=1/p_{d}$ and this result
remains valid when, e. g., we minimize the ratio of standard deviations
$\sqrt{Var\left[Y'\right]/N^{-1}}$ instead of $Var\left[Y'\right]/N^{-1}$.
When errors are not totally destructive, the error will still accumulate
in the random-walk-like fashion and will eventually become totally
destructive. In this case, we may redefine $p_{d}$ accordingly (which
we do not do in this paper) so that, on average, it takes approximately
$1/p_{d}$ single electron processes for the accumulated error to
be fully destructive.

Finally, we remark on relativistic effects. Since the TEM we have
in our mind may well operate at voltages as high as 300 keV, there
will be relativistic effects. In the rest of the paper we use the
so-called relativistically corrected Schrodinger equation, which is
a good approximation to the Dirac equation in our case \cite{relativistic scattering}.
The equation \emph{is} the standard time-independent Schrodinger equation,
but the mass $m$ and the energy $E$ are interpreted respectively
as $m=\gamma m_{0}$ and $E=\left(E_{r}^{2}-m_{0}^{2}c^{4}\right)/2\gamma m_{0}c^{2}$,
where $\gamma=\left(1-\beta^{2}\right)^{-1/2}$, $\beta=v/c$, $m_{0}$
is the electron rest mass, and $E_{r}$ is the relativistic electron
energy including the rest energy $m_{0}c^{2}$.

\section{Errors due to coherent processes\label{sec:Coherent errors}}

In this section, we study the extent to which the similar intensity
map condition $\left|a_{j}\right|\cong\left|b_{j}\right|$ is satisfied,
for two cases of the probe electron beam geometry.

\subsection{Focused incident beam\label{sub:Focussed-incident-beam}}

First we study a simple case, in which the two regions $S_{0}$ and
$S_{1}$ are adjacent two small regions of the same size of $\sim1\mathrm{nm}$.
We consider a $\sim30\mathrm{nm}$ thick specimen. We assume that
the electrons are detected on a diffraction plane. Then the problem
is effectively about the diffraction pattern from a small spot and
the question is how smooth the intensity on the detector is. Let
us focus on the region $S_{0}$. Let the $z$-axis be parallel to
the direction of the incident wave propagation and let the position
of the specimen be at $z\simeq0$. Atoms in the specimen are labeled
by an integer $s$ and are respectively located at $\boldsymbol{r}_{s}=\left(x_{s},y_{s},z_{s}\right)$.
Let $f_{s}$ be the scattering amplitude of the $s$-th atom corresponding
to its elemental identity. We assume the electron waves to be monochromatic
and write the wavenumber $k$. We ignore multiple scatterings in the
specimen. We want the incident beam to be Gaussian with the waist
size $w_{0}$ and assume that the diffraction-limited waist of the
Gaussian beam is on the specimen. Without loss of generality, we compute
the wave amplitude at a point P in the far field on the $xz$-plane,
where $z=z_{L}$ is large and $x$ coordinate is given by $\tan\theta=x/z_{L}$.
Far from the specimen, the transmitted Gaussian wave has the known
form in the paraxial approximation (See Appendix A) \begin{equation}
\psi_{T}=\frac{-ikw_{0}^{2}e^{ikr}}{2z}e^{-\frac{k^{2}w_{0}^{2}}{4}\left(\frac{x}{z}\right)^{2}}\label{eq:transmitted wave far field}\end{equation}
that can be approximated as\begin{equation}
\psi_{T}=\frac{-ikw_{0}^{2}e^{ikr}}{2r}e^{-\frac{k^{2}w_{0}^{2}\theta^{2}}{4}}\label{eq:transmitted wave far field 2}\end{equation}
Consider a single elastic scattering event. The amplitude of the scattered
wave at the point P is found to be (See Appendix A) \begin{equation}
\psi_{P}=\frac{e^{ikr}}{r}\sum_{s}e^{-\frac{x_{s}^{2}+y_{s}^{2}}{w_{0}^{2}}}f_{s}\left(\theta\right)e^{-ikx_{s}\theta}\label{eq:scattered wave 1}\end{equation}
For simplicity, let us first consider the contributions to the wave
amplitude $\psi_{P}$ only from \emph{uniformly} distributed carbon
atoms, which we write $\psi_{P,C}$. The scattering amplitude $f_{s}$
is that of carbon $f_{C}$. Let the number density of carbon atoms
per unit area perpendicular to the optical axis be $n_{C}$. The value
of $\psi_{P,C}$ is\[
\psi_{P,C}=\frac{e^{ikr}}{r}f_{C}\left(\theta\right)n_{C}\iint e^{-\frac{x^{2}+y^{2}}{w_{0}^{2}}}e^{-ikx\theta}dxdy\]
 \begin{equation}
=\pi f_{C}\left(\theta\right)n_{C}w_{0}^{2}\frac{e^{ikr}}{r}e^{-\frac{k^{2}w_{0}^{2}}{4}\theta^{2}}\label{eq:scattered wave 1C}\end{equation}
Simple addition will extend this result to the case of multiple chemical
elements. Comparing with equation (\ref{eq:transmitted wave far field 2}),
we see that the scattered wave $\psi_{P,C}$ does almost nothing more
than shifting the phase of the transmitted wave, especially when the
angular spread of $f_{C}\left(\theta\right)\sim10\mathrm{mrad}$ is
larger than the Gaussian incident beam spread, which is about a few
mrad in the present case.

It is the random, as opposed to uniform, arrangement of atoms that
we need to analyze to evaluate how well the similar intensity map
condition is fulfilled. Note that $z$-coordinates $z_{s}$ of atoms
do not appear in equation (\ref{eq:scattered wave 1}) because of
approximations discussed in Appendix A. Hence, what matters is only
the projected atomic distribution of the specimen to a plane perpendicular
to the optical axis. Henceforth we assume that such a projected atomic
distribution is completely random. This assumption obviously fails
if, for example, the specimen is atomically thin because interatomic
distance cannot be too small in real specimens, and there may well
be other objections to this assumption. However, we assume that this
assumption will give us a good guide especially when the specimen
is sufficiently thick. 

Computation of equation (\ref{eq:scattered wave 1}) beyond the average
case resembles the analysis of random walk. Let the specimen plane
be divided into many thin concentric rings centered at the optical
axis with the radius $r=\sqrt{x^{2}+y^{2}}$ and small width $\Delta r$.
Label the rings with a natural number $j$ in the order of increasing
$r$. Assume that, despite the small width $\Delta r$, each ring
still contains many atoms. Again for simplicity, we consider only
carbon atoms. Let the number of carbon atoms in the $j$-th ring be
$n_{j,C}$. Equation (\ref{eq:scattered wave 1}) is then rewritten
as \begin{equation}
\psi_{P,C}=\frac{e^{ikr}}{r}f_{C}\left(\theta\right)\sum_{j}\left(e^{-\frac{r_{j}^{2}}{w_{0}^{2}}}\sum_{s\epsilon R_{j,C}}e^{-ikx_{s}\theta}\right)\end{equation}
where $R_{j,C}$ is a set of integers consisting of the indices of
carbon atoms that belong to the $j$-th ring. Hence the cardinality
of $R_{j,C}$ is $n_{j,C}$. Let $\theta$ be large enough so that
all sort of phase angles in the range $[0,2\pi)$ appear in the sum
$\sum_{s\epsilon R_{j,C}}e^{-ikx_{s}\vartheta}$. In other words,
we consider the region outside the peak of transmitted wave in the
diffraction pattern. In this case, we may regard the sum as 2-dimensional
random walk on the Gaussian plane and we have $\sum_{s\epsilon R_{j,C}}e^{-ikx_{s}\theta}\cong\sqrt{n_{j,C}}e^{i\phi_{j}}$,
where $\phi_{j}$ is a random phase. More precisely, we accept that
the angle $-kx_{s}\theta$ behaves as a random variable with the uniform
distribution because of the large $\theta$ and the supposedly random
arrangement of the carbon atoms from location to location within the
specimen. The phase factors $e^{-ikx_{s}\theta}$ are mutually independent
random variables and $\psi_{P,C}$ is also a random variable in this
view. Clearly, the expectation value $E\left\langle \psi_{P,C}\right\rangle $
is zero because of the random phase, which is consistent with what
equation (\ref{eq:scattered wave 1C}) states. (Note that the expectation
value here is with respect to the random arrangement of atoms and
has nothing to do with the randomness in quantum measurement.) However,
the expectation value of the absolute square of the wavefunction is
nonzero because multiplication of phase factors $\exp\left(ikx_{s}\theta\right)\cdot\exp\left(-ikx_{s'}\theta\right)$
results in $1$ when $s=s'$:\[
E\left\langle \left|\psi_{P,C}\right|^{2}\right\rangle =\left[\frac{f_{C}\left(\theta\right)}{r}\right]^{2}\sum_{j}n_{j,C}e^{-\frac{2r_{j}^{2}}{w_{0}^{2}}}=\left[\frac{f_{C}\left(\theta\right)}{r}\right]^{2}\sum_{j}n_{C}e^{-\frac{2r_{j}^{2}}{w_{0}^{2}}}2\pi r_{j}\Delta r_{j}\]
\begin{equation}
\cong\left[\frac{f_{C}\left(\theta\right)}{r}\right]^{2}\int_{0}^{\infty}n_{C}e^{-\frac{2r^{2}}{w_{0}^{2}}}2\pi rdr=\frac{\pi}{2}n_{C}w_{0}^{2}\left[\frac{f_{C}\left(\theta\right)}{r}\right]^{2}\end{equation}
Since we assume that the atomic arrangements are independent to each
other among the chemical elements, the contributions from elements
other than carbon can be accounted for by simple addition to the above
expression. Crucially, the variance of $\left|\psi_{P,C}\right|^{2}$,
that is $Var\left\langle \left|\psi_{P,C}\right|^{2}\right\rangle =E\left\langle \left|\psi_{P,C}\right|^{4}\right\rangle -\left\{ E\left\langle \left|\psi_{P,C}\right|^{2}\right\rangle \right\} ^{2}$,
equals $\left\{ E\left\langle \left|\psi_{P,C}\right|^{2}\right\rangle \right\} ^{2}$.
This may be verified by noting that multiplication of four phase factors
$\exp\left(ikx_{s}\vartheta\right)\cdot\exp\left(-ikx_{s'}\vartheta\right)\cdot\exp\left(ikx_{s''}\vartheta\right)\cdot\exp\left(-ikx_{s'''}\vartheta\right)$
results in $1$ when either\emph{ }\{$s=s'$ and $s''=s'''$\}, or\emph{
}\{$s=s'''$ and $s'=s''$\} holds. (We ignore the relatively rare
case of $s=s'=s''=s'''$.) Hence the diffraction pattern intensity
fluctuates as much as its own average. (The physical manifestation
of it is known as the speckle pattern.) This in turn means that the
similar intensity map condition $\left|a_{j}\right|\cong\left|b_{j}\right|$
is \emph{not} satisfied whenever an electron is detected outside the
central peak described in equation (\ref{eq:transmitted wave far field 2}).
Since typical angular spread associated with an elastic electron scattering
$\cong10\mathrm{mrad}$ is much larger than the spread of the incident
beam $\cong1.3\mathrm{mrad}$ (See Appendix A), essentially every
elastic scattering process fails to satisfy the similar intensity
map condition, and hence entails destruction of the CPB quantum state.
This suggests that the ordinarily welcome high-angle elastic scattering
processes do some harm here, because it provides unwanted high-resolution
information that generates the uncontrollable speckle pattern. This
is a rather curious feature of quantum measurement in this particular
case, but the feature may be more generic. If so, we might state that
we generally do not want to know more than needed in quantum measurement.

To compute the frequency of elastic scattering, it suffices to know
the total elastic scattering cross sections for each relevant chemical
element\cite{NIST database}, the number of atoms per unit area for
these elements (See Appendix B), and the specimen thickness that we
assume to be $30\mathrm{nm}$. It is found that $300\mathrm{keV}$
electrons are elastically scattered with the probability $p_{d}=5.2\%$.
If inelastic processes \emph{were} negligible, according to the argument
in Sec. \ref{sub:Several-remarks}, this means that we can use $k_{m}=19.4$
electrons on average in a k-electron process, which would results
in a contrast enhancement by a rather modest factor $\sqrt{k_{m}/e}\cong2.7$.
Moreover, inelastic processes of course are not negligible and we
will deal with them later.

\subsection{Diverging incident beam\label{sub:Diverging-incident-beam}}

We now turn to the second case, where a circular region $S_{0}$ surrounds
a smaller circular region $S_{1}$. These two regions share the central
point. This configuration appeared in a simulation study previously
reported \cite{ent-assisted EM}. To be specific, we consider the
following situation: The incident beam has a relatively large divergence
angle $\theta_{G}'=2/kw_{0}'$ (Notice that we add primes to these
variables in the present diverging beam case), which is larger than
the characteristic angle $\cong10\mathrm{mrad}$ of the scattering
amplitude functions $f_{s}\left(\theta\right)$. For $300\mathrm{keV}$
electrons, the diffraction-limited waist size $w_{0}'$ of the Gaussian
beam with the divergence angle $\theta_{G}'=40\mathrm{mrad}$ is $w_{0}'=16\mathrm{pm}$,
which appears to be feasible, considering there already is a $<50\mathrm{pm}$
probe \cite{50 pm probe}. We set the distance $\Delta z$ between
the incident beam waist and the \emph{central plane}, which is at
the midpoint between the entrance and exit surfaces of the $30\mathrm{nm}$-thick
specimen, to be $52.5\mathrm{nm}$ or $22.5\mathrm{nm}$ for the regions
$S_{0}$ and $S_{1}$, respectively. Hence, on the entrance surface,
the beam diameter is $3.0\mathrm{nm}$ and $0.6\mathrm{nm}$ for the
regions $S_{0}$ and $S_{1}$, respectively. The beam size increases
to $5.4\mathrm{nm}$ and $3.0\mathrm{nm}$ respectively for the cases
$S_{0}$ and $S_{1}$ at the exit surface of the specimen. This beam
spread is large, but we must accept this because of plasmon scattering
discussed in Sec. \ref{sec:Effect-of-inelastic}. (Suitably designed
data processing may solve this large-beam-spread problem. However,
it goes beyond the scope of this paper.) Again, the electrons are
detected on a diffraction plane. We define a parameter $\varepsilon\equiv k\left(w'_{0}\right)^{2}/2\Delta z$,
which turns out to be small ($\sim0.02$) in our case (Appendix A).
While for transmitted wave we can use equation (\ref{eq:transmitted wave far field 2})
with $w_{0}$ replaced with $w_{0}'$, the amplitude of the scattered
wave in the present case is found to be (See Appendix A)\begin{equation}
\psi'_{P,C}=\frac{-i\varepsilon e^{ikr}}{r}\sum_{s}e^{i\frac{k}{2\Delta z}\left[\left(x_{s}-\Delta z\theta\right)^{2}+y_{s}^{2}\right]}e^{-\varepsilon\frac{k}{2\Delta z}\left(x_{s}^{2}+y_{s}^{2}\right)}f_{s}\left(\theta-\frac{x_{s}}{\Delta z}\right)\label{eq:scattered wave 2}\end{equation}
where $\theta$ again is the scattering angle towards $+x$ direction.

To evaluate equation (\ref{eq:scattered wave 2}), we again focus
on carbon atoms and first assume that these atoms are uniformly distributed.
Then the above is\[
\psi'_{P,C}\cong\frac{-i\varepsilon e^{ikr}}{r}n_{C}\int\int dxdyf_{C}\left(\frac{-1}{\Delta z}\left(x-\Delta z\theta\right)\right)e^{i\frac{k}{2\Delta z}\left[\left(x-\Delta z\theta\right)^{2}+y^{2}\right]}e^{-\varepsilon\frac{k}{2\Delta z}\left(x^{2}+y^{2}\right)}\]
\begin{equation}
=\frac{-i\varepsilon e^{ikr}}{r}n_{C}\sqrt{\frac{2\pi i\Delta z}{k\left(1+i\varepsilon\right)}}\int dxf_{C}\left(\left(-1/\Delta z\right)\left(x-\Delta z\theta\right)\right)e^{i\frac{k}{2\Delta z}\left(x-\Delta z\theta\right)^{2}}e^{-\varepsilon\frac{k}{2\Delta z}x^{2}}\label{eq:uniform scattered wave 2}\end{equation}
The integral is a convolution of two functions $f_{C}\left(x/\Delta z\right)e^{i\frac{k}{2\Delta z}x^{2}}$
and $e^{-\varepsilon\frac{k}{2\Delta z}x^{2}}$. The former has two
factors. The factor $f_{C}\left(-x/\Delta z\right)$ quickly goes
to zero above $\left|x/\Delta z\right|\sim10\mathrm{mrad}$ and the
other factor $e^{i\frac{k}{2\Delta z}x^{2}}$ rapidly oscillates also
above $\left|x/\Delta z\right|\sim\sqrt{4\pi/k\Delta z}=10\mathrm{mrad}$
(See Appendix A). The latter function $e^{-\varepsilon\frac{k}{2\Delta z}x^{2}}$
has a broader profile than the former when plotted against $x/\Delta z$,
by a factor $1/\sqrt{\varepsilon}\cong7$.

A crude but useful approximation to equation (\ref{eq:uniform scattered wave 2})
is to replace $f_{C}\left(\left(-1/\Delta z\right)\left(x-\Delta z\vartheta\right)\right)$
with $f_{C}\left(0\right)$ because unless the argument of the function
$f_{C}$ is close to zero, the factor $e^{i\frac{k}{2\Delta z}\left(x-\Delta z\vartheta\right)^{2}}$
oscillates anyway, making its contribution to the integral small.
(We neglected the fact that the 'width' of $f_{C}$ is not so wide.
This is why the approximation is crude.) Once we accept this, the
integration in equation (\ref{eq:uniform scattered wave 2}) can be
carried out, e.g. by noting that it is a convolution of two Gaussian
functions, to obtain\begin{equation}
\psi'_{P,C}\cong n_{C}\pi\left(w_{0}'\right)^{2}\cdot\frac{f_{C}\left(0\right)e^{ikr}}{r}\cdot\frac{1}{1+i\varepsilon}e^{-\left(\frac{\theta}{\theta'_{G}}\right)^{2}\frac{1}{1+i\varepsilon}}\label{eq:uniform scattered wave 3}\end{equation}
or equivalently\begin{equation}
=-i\frac{e^{ikr}k\left(w'_{0}\right)^{2}}{2r}\left(if_{C}\left(0\right)\lambda n_{C}\right)\frac{1}{1+i\varepsilon}e^{-\frac{k^{2}\left(w'_{0}\right)^{2}\theta^{2}}{4}\cdot\frac{1}{1+i\varepsilon}}\label{eq:uniform scattered wave 4}\end{equation}
where $\theta'_{G}=2/kw'_{0}$ is the half angle of the incident wave.
Comparing with equation (\ref{eq:transmitted wave far field 2}),
if we take the two $1/\left(1+i\varepsilon\right)$ factors as close
to $1$, then equation (\ref{eq:uniform scattered wave 4}) represents
a phase shift of the incident wave by an angle $f_{C}\left(0\right)\lambda n_{C}$,
as long as this quantity is small. This finding is consistent with
the result on weak phase object described in Appendix A of Ref. \cite{Okamoto-3}.

Again, in order to evaluate the validity of the similar intensity
map condition, random atomic distributions must be considered instead
of the uniform distribution. In this case, the scattered wave spreads
with the half angle that roughly equals addition of $\theta'_{G}$
and the characteristic angle of elastic scattering $\theta_{0}$.
(In the above uniform atomic distribution case, the half angle does
not exceed $\theta'_{G}$ because beyond this angle, scattered waves
from the atoms destructively interfere.) 

Let us consider two angular regions separately. First, if the angle
$\theta$ is less than $\theta'_{G}$, the transmission wave $\psi'_{T}$
is the largest component that goes into this angular region. The scattered
waves from each atom, on the other hand, constructively interfere
more or less, because of the small $\theta$. The resultant $\psi'_{P,C}$,
with $\pi/2$ phase shift, adds to $\psi'_{T}$. As noted above, $\psi'_{T}+\psi'_{P,C}$
is a phase-shifted (by the angle $f_{C}\left(0\right)\lambda n_{C}$)
version of the incident wave. If the atomic density $n_{C}$ varies
due to randomness, then the phase shift also varies. However, this
'randomness' represents exactly what we want to measure. The phase
shift $f_{C}\left(0\right)\lambda n_{C}$ may not be small by itself,
but the difference of two phase shifts $\Delta\varphi$ between the
regions $S_{0}$ and $S_{1}$ are small. Indeed, at the $\mathrm{nm}$
scale, $\Delta\varphi$ is expected to be $\sim5\mathrm{mrad}$ for
$300\mathrm{keV}$ electrons \cite{ent-assisted EM,Okamoto-3} and
what this phase difference is expected to cause on the wave amplitude
should be equal to or higher than the second-order correction, which
roughly is $\sim\Delta\varphi^{2}\sim10^{-4}$. (In order to exactly
compute the second order correction, we must go beyond the first-order
perturbation theory, which we do not do in the present work.) Hence
in this region of $\theta<\theta'_{G}$, we do not expect problems
regarding the similar intensity map condition.

Second, consider the outer region $\theta'_{G}<\theta<\theta'_{G}+\theta_{0}$.
Here, we expect the 'speckle pattern' and hence we assume that the
scattered waves from each atom add with random phases. Hence, the
average intensity in this region is the sum of the intensities of
each atom:\[
E\left\langle \left|\psi'_{P,C}\right|^{2}\right\rangle =\left(\frac{\varepsilon}{r}\right)^{2}\sum_{s}e^{-\varepsilon\frac{k}{\Delta z}\left(x_{s}^{2}+y_{s}^{2}\right)}f_{C}^{2}\left(\theta-\frac{x_{s}}{\Delta z}\right)\]
\[
\cong n_{C}\left(\frac{\varepsilon}{r}\right)^{2}\int\int dxdye^{-\varepsilon\frac{k}{\Delta z}\left(x^{2}+y^{2}\right)}f_{C}^{2}\left(\theta-\frac{x}{\Delta z}\right)\]
\begin{equation}
=\sqrt{\frac{\pi}{2}}n_{C}\frac{\varepsilon w'_{0}}{r^{2}}\int dxe^{-\varepsilon\frac{kx^{2}}{\Delta z}}f_{C}^{2}\left(\theta-\frac{x}{\Delta z}\right)\end{equation}
Or equivalently, \begin{equation}
=\sqrt{\frac{\pi}{2}}\frac{k\left(w'_{0}\right)^{3}}{2r^{2}}N_{C}\int d\gamma e^{-2\left(\frac{\gamma}{\theta'_{G}}\right)^{2}}g_{C}^{2}\left(\theta-\gamma\right)\equiv\sqrt{\frac{\pi}{2}}\frac{k\left(w'_{0}\right)^{3}}{2r^{2}}N_{C}G_{C}\left(\theta\right)\end{equation}
where $g_{C}=f_{C}/a_{0}$ is the scattering amplitude normalized
by the Bohr radius $a_{0}$ that is provided in Ref. \cite{NIST database}
and $N_{C}\equiv n_{C}a_{0}^{2}$ is the number of carbon atoms in
the area $a_{0}^{2}$. This expression may easily be extended to the
multiple chemical element case by simple addition. We get the standard
deviation of the amplitude $\psi'_{P}$, \begin{equation}
SD\left\langle \psi'_{P}\right\rangle =\sqrt{E\left\langle \left|\psi'_{P}\right|^{2}\right\rangle }=\sqrt{E\left\langle \left|\psi'_{P,C}\right|^{2}\right\rangle +E\left\langle \left|\psi'_{P,N}\right|^{2}\right\rangle +E\left\langle \left|\psi'_{P,O}\right|^{2}\right\rangle +\cdots}\end{equation}
which provides essentially the expected amplitude, but this amplitude
should fluctuate much, as in the focused incident beam case described
in Sec. \ref{sub:Focussed-incident-beam}. Hence if we detect an electron
in the region $\theta'_{G}<\theta<\theta'_{G}+\theta_{0}$, then the
measurement would be spoiled because the similar intensity map condition
is not valid in this region. More precisely, at angles $\theta$ where
$\left|\psi'_{T}\right|\leq SD\left\langle \psi'_{P}\right\rangle $,
we expect that the similar intensity map condition is not satisfied.
The condition is written down explicitly as\begin{equation}
e^{-2\left(\frac{\theta}{\theta'_{G}}\right)^{2}}\leq\sqrt{\frac{\pi}{2}}\theta'_{G}H\left(\theta\right)\label{eq:condition}\end{equation}
where $H\left(\theta\right)\equiv N_{C}G_{C}\left(\theta\right)+N_{N}G_{N}\left(\theta\right)+N_{O}G_{O}\left(\theta\right)+\cdots$.
Let the smallest $\theta$ that satisfies equation (\ref{eq:condition})
be $\theta_{c}$. The function $H\left(\theta\right)$ can be evaluated
numerically using published data \cite{NIST database} for elements
H, C, N, O, S, which results in $\theta_{c}=71.9\mathrm{mrad}$.

Finally, we compute the probability $p_{d}'$ of undesirable elastic
scattering that satisfies the condition (\ref{eq:condition}). This
can be expressed as the ratio between the integrated scattered wave
intensity into undesirable angular region and the integrated incident
wave intensity.\begin{equation}
p'_{d}=\frac{\int_{\theta_{c}}^{\pi}E\left\langle \left|\psi'_{P}\right|^{2}\right\rangle 2\pi\sin\theta d\theta}{\int_{0}^{\pi}\left|\psi'_{T}\right|^{2}2\pi\sin\theta d\theta}\end{equation}
From equation (\ref{eq:transmitted wave far field 2}), the denominator
is computed as\begin{equation}
\int_{0}^{\pi}\left|\psi'_{T}\right|^{2}2\pi\sin\theta d\theta\cong2\pi\int_{0}^{\infty}\left|\psi'_{T}\right|^{2}\theta d\theta=\frac{\pi\left(w'_{0}\right)^{2}}{2r^{2}}\end{equation}
 The numerator involving $H\left(\vartheta\right)$ should be handled
numerically. We obtain a much more favorable figure than the focused
incident beam case\begin{equation}
p'_{d}=\frac{\sqrt{8\pi}}{\theta'_{G}}\int_{\theta_{c}}^{\pi}H\left(\theta\right)\sin\theta d\theta=4.5\times10^{-3}\end{equation}
The inverse of this is $k_{m}=1/p_{d}'\cong222$. This is more than
sufficient because it would make $k_{m}\Delta\varphi>2\pi$ for large
values of $\Delta\varphi$.

It should be noted that the above computation is little more than
order estimation because we simply asserted that $\left|\psi'_{T}\right|\leq SD\left\langle \psi'_{P}\right\rangle $
means complete spoiling of the measurement, and have not addressed
the intermediate-level damaging of the CPB state etc. However, complete
characterization of such processes would demand full computer simulations
and is beyond the scope of the present paper.

\section{Effect of inelastic scattering\label{sec:Effect-of-inelastic}}

\subsection{A brief review}

We begin by reviewing several known facts, many of them found in \cite{Egerton textbook},
about inelastic electron scattering, which have been established through
decades of theoretical and experimental studies. Inelastic scattering
by definition involves excitation of electrons within the specimen,
and is \emph{not} characterized by energy loss \emph{per se}. (In
fact, the probe electrons do lose a small amount of energy upon elastic
scattering, either by high-angle scattering or generation of phonons.)
Conversely, elastic scattering processes are simply ones that are
not inelastic scattering, and is not characterized by the ability
to produce interference fringes. In fact, interference has been observed
in experiments on inelastic scattering \cite{Kimoto & Matsui}.

Inelastic scattering processes roughly divide into excitations of
outer-shell electrons and the inner-shell counterpart. Although inner
shell excitation processes are generally important in EELS because
it carries elemental information, the associated cross sections are
small and these are insignificant in the present context, as will
be shown later. The outer shell excitations, on the other hand, are
more complex because the states of valence electrons depends on the
chemical environment of the atom under study. These excitations are
typically collective in nature and in many cases, including the case
of insulators and polymers, these are treated as plasmons with energy
loss $E\sim20\mathrm{eV}$ \cite{polymer radiation damage}. Because
of the dependence on chemical bonding to neighboring atoms, inelastic
scattering cross sections are difficult to predict with certainty
\cite{Langmore & Smith}. Despite such uncertainty, it is generally
accepted that energy loss at around $E\sim20\mathrm{eV}$ does represent
excitation of plasmons in vitrified biological specimens, while there
are other relatively minor structures with smaller $E$ in the energy
loss spectrum, which is thought to represent localized excitons \cite{Leapman & Sun}.

Here are several additional facts about inelastic scattering. First,
inelastic scattering experimentally has narrower scattering angle
spread when compared to elastic scattering \cite{Langmore & Smith}.
As will be explained shortly, however, it is not easy to talk about
a characteristic or average angle associated with it. Second, inelastic
scattering happens roughly twice as often as elastic scattering in
organic materials \cite{e-crystallography book,Wall Isaacson & Langmore}.
Third, inelastic scattering comes from long-range Coulombic interaction
between the probe electron and a bound electron of an atom. Hence
inelastic scattering can happen even when the electron beam does not
directly hit the atom \cite{polymer delocalization}. This is closely
related to the concept of \emph{delocalization} \cite{delocalization egerton}. 

Bethe theory provides a double-differential inelastic scattering cross
section with energy loss $E$ \cite{Egerton textbook}. It has the
following dependence on the scattering angle $\theta$:\begin{equation}
\frac{d^{2}\sigma}{d\Omega dE}\cong\frac{8a_{0}^{2}R^{2}}{Em_{0}v^{2}}\left(\frac{1}{\theta^{2}+\theta_{E}^{2}}\right)\frac{df}{dE}\label{eq:bethe formula}\end{equation}
where $a_{0}$ is the Bohr radius, $R=13.6\mathrm{eV}$ is the Rydberg
energy, $m_{0}$ is the electron rest mass, $v$ is velocity of the
incident electrons, and $f$ is what is called the generalized oscillator
strength (GOS). The GOS is modified when a solid specimen is considered
because of factors such as chemical bonding. However, what is modified
is the energy dependence of the GOS and the angular distribution of
inelastic scattering is mostly robust \cite{Egerton textbook}. We
evaluate equation (\ref{eq:bethe formula}) at the typical plasmon
energy loss $E=20\mathrm{eV}$. The angle $\theta_{E}$ is given as
$ $ \begin{equation}
\theta_{E}=\frac{E}{mv^{2}}\label{eq:theta-E}\end{equation}
where $m$ is the relativistic electron mass. For $300\mathrm{keV}$
electrons and $E=20\mathrm{eV}$, $\theta_{E}\cong41\mathrm{\mu rad}$.
Let $k_{0}$ be the wavenumber of the incident electrons. It is tempting
to identify $b_{max}\equiv1/k_{0}\theta_{E}\cong7.6\mathrm{nm}$ with
the delocalization length of inelastic scattering, which should give
a limit of resolution when inelastically scattered electrons are used
for imaging. However, the distribution $\left(\theta^{2}+\theta_{E}^{2}\right)^{-1}$
has a long tail. Consequently, identification of $b_{max}$ with the
delocalization length has led to some confusion in the past because
the apparent experimental resolution is much better than $b_{max}$
\cite{delocalization muller-silcox}.

Equation (\ref{eq:bethe formula}) allows us to compute the intensity
pattern of inelastically scattered electrons at the diffraction plane.
From this, one can infer the wavefunction $\psi_{inel}\left(r\right)$,
where $r$ is the distance from the atom that scatters the electron,
of the inelastically scattered electron right after the scattering
by assuming that $\psi_{inel}\left(r\right)$ is real. This is done
by ignoring the $\theta$ dependence of the GOS and Fourier transform
the square root of the diffraction intensity pattern to get the wavefunction
back at the specimen plane. One obtains \cite{Egerton textbook}\begin{equation}
\psi{}_{inel}\left(r\right)\propto\frac{e^{-r/b_{max}}}{r}\label{eq:sqrt PSF}\end{equation}

\subsection{Inelastic scattering in entanglement-assisted electron microscopy}

\subsubsection{Localized inelastic scattering}

We first examine, at a relatively abstract level, how inelastic scattering
affects entanglement-assisted electron microscopy. We saw in Sec.
\ref{sub:A-brief-review} that the incident electron to the specimen
is entangled with the CPB as equation (\ref{eq:pre whole state})
indicates. We first simplify the situation and suppose that the electron
beam excites a single atom in the specimen. In other words, we consider
a localized excitation, such as excitation of an inner-shell electron.
Let the ground state and excited states of the atom be $|g>_{a},|e_{1}>_{a},|e_{2}>_{a},\cdots$
, where the subscript '\emph{a}' stands for the word 'atom'. The initial
state before interaction between the probe electron and the specimen
is \begin{equation}
|\Phi_{1}>=\left(c_{0}|0>_{b}\otimes|0>+c_{1}|1>_{b}\otimes|1>\right)\otimes|g>_{a}\label{eq:delocalization principle1}\end{equation}
where we use the letter $\Phi$ instead of $\Psi$ (see e. g. equation(\ref{eq:pre whole state}))
to indicate that the atom is now included to the system under consideration.
After collision between the probe electron and the specimen, probability
amplitudes that correspond to the inelastic scattering develop, although
these may be of relatively small amplitude. The wave function for
the entire system is written as\[
|\Phi_{2}>=c_{0}|0>_{b}\otimes\left(a_{00}|0_{g}>\otimes|g>_{a}+a_{01}|0_{e1}>\otimes|e_{1}>_{a}+a_{02}|0_{e2}>\otimes|e_{2}>_{a}+\cdots\right)\]
\begin{equation}
+c_{1}|1>_{b}\otimes\left(a_{10}|1_{g}>\otimes|g>_{a}+a_{11}|1_{e1}>\otimes|e_{1}>_{a}+a_{12}|1_{e2}>\otimes|e_{2}>_{a}+\cdots\right)\label{eq:delocalization principle2}\end{equation}
where $|0_{g}>,|1_{g}>$ represents zero-energy-loss component of
the scattered electron wave and $|0_{e1}>,|1_{e1}>$ etc. are inelastically
scattered electron waves. Let these states be properly normalized,
i. e. $<1_{e1}|1_{e1}>=1$ etc. Then, for example, the coefficient
$a_{02}$ is a probability amplitude for the probe electron going
through the region $S_{0}$ to excite the atom to the second excited
state. Now, suppose that the state of the excited atom is found to
be, for example, $|e_{1}>_{a}$ by a projective measurement on the
atom. This corresponds to the situation where inelastic scattering
actually happens. (We of course do not measure the state of the atom
experimentally, but we represent the whole complex processes after
the inelastic scattering that involves thermalization, decoherence
etc. by a single projective measurement where the 'measurement outcome'
is hidden to the experimenter.) Then the system consisting of the
CPB and probe electron is left in the state \begin{equation}
|\Psi_{2}'>=c_{0}a_{01}|0>_{b}\otimes|0_{e1}>+c_{1}a_{11}|1>_{b}\otimes|1_{e1}>\label{eq:delocalization principle3}\end{equation}

We are in trouble here for the following reasons. The electron states
$|0_{e1}>,|1_{e1}>$ are rather different from $|0_{g}>,|1_{g}>$
as shown in Fig. 2. If expressed as wavefunctions, for example, the
state $|0_{g}>$ is a diverging Gaussian beam as discussed in Sec.
\ref{sub:Diverging-incident-beam}. On the other hand, recall that
the inelastically scattered electron wave generates the diffraction
pattern consistent with equation (\ref{eq:bethe formula}) and corresponds
to the wavefunction (\ref{eq:sqrt PSF}), \emph{if} a plane incident
wave is employed. Because of the linearity of quantum mechanics, the
wavefunction corresponding to the state $|0_{e1}>$ equals the product
of the Gaussian incident wavefunction and the equation (\ref{eq:sqrt PSF}).
First, such a product may be vanishing if the overlap between the
Gaussian envelope and the function (\ref{eq:sqrt PSF}) is small.
In this case, only one term of equation (\ref{eq:delocalization principle3})
survives and we get amplitude imbarance, or an amplitude error. Second,
the phase compensation step described in Sec. \ref{sub:A-brief-review}
may fail because the optical path length difference is no longer the
difference of distances between the detection pixel and the two regions
$S_{0}$ and $S_{1}$ respectively. Rather, this quantity now involves
the \emph{modified} two regions $S'_{0},S'_{1}$ because the form
of $|1_{e1}>$ is localized at around the scattering atom and is different
from $|1_{g}>$. Since we do not know the exact position of the inelastically
scattering atom, we cannot compensate for this modification when performing
the phase compensation step. Hence, we have problems in terms of both
the amplitude and phase.

\begin{figure}
\begin{center}
\includegraphics[width=0.45\textwidth]{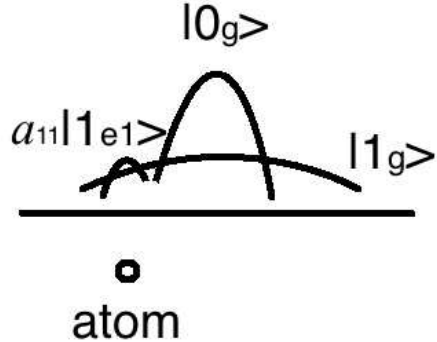}

\caption{Schematic figure of wavefunctions on the specimen plane. The position
of the atom is shown, where inelastic scattering occurs. The state
$|0_{e1}>$ is not shown because its amplitude is too small, due to
the fact that the atom is outside the spatial extent of the state
$|0_{g}>$.}

\end{center}
\end{figure}

Since the inner-shell excitations are problematic for the above reasons,
we have used the SIGMAK3 and SIGMAL3 programs \cite{egerton programs MATLAB}
to evaluate the total K-shell and L-shell cross sections for $300\mathrm{keV}$
electrons for the elements C, N, O, and S. (Hydrogen does not have
an inner shell.) We combined the result with data presented in Appendix
B, which can easily be translated to atomic number density-per-area
data for a typical $30\mathrm{nm}$-thick biological specimen. We
find that the probability for a single incident electron to cause
K-shell or L-shell ionization in such $30\mathrm{nm}$-thick specimen
is $8.6\times10^{-4}$. This figure is small enough to perform entanglement
assisted electron microscopy to our advantage.

\subsubsection{Delocalized inelastic scattering}

The most dominant outer-shell excitations are usually plasmons and
here we focus on inelastic scattering by plasmons. Let us first develop
a simple but admittedly inaccurate model to grasp essential physics
involved. The purpose of doing so is not to estimate quantities we
want to know, but to make a connection between the localized case
mentioned above to the case of delocalized scattering involving plasmons. 

We are interested in specimens consisting of biological molecules
embedded in vitreous ice. Clearly, it is not adequate to model such
an insulating system by free electrons in a box, as done in studies
of plasmons in metals. Without further justifications, let us accept
that essential physics of such an insulating system is captured by
a set of $N$ 1-dimensional harmonic oscillators, which are arranged
on a line with a constant interval $a$. (The actual specimen is more
like 2-dimensional, but we can oversimplify as long as the purpose
is only to grasp essential physics.) Each harmonic oscillator models
an atom with a single oscillating valence electron and the immobile
rest. Assume that all neighboring two electrons interact as if these
are connected by a mechanical spring. This assumption is hard to justify,
but for the sake of simplicity we proceed. Then the model is similar
to the 1-dimensional model for phonons. The hamiltonian is\begin{equation}
H=\sum_{s=1}^{N}\left\{ \frac{p_{s}^{2}}{2m_{0}}+\frac{C}{2}\left(q_{s+1}-q_{s}\right)^{2}+\frac{C'}{2}q_{s}^{2}\right\} \label{eq:hamiltonian}\end{equation}
where $p_{s}$ and $q_{s}$ are the momentum and position of the electron
at the $s$-th site, $m_{0}$ is the electron mass, $C$ and $C'$
are 'spring constants' of interaction between the $s$-th electron
and the neighboring $s+1$-th electron; and interaction between the
$s$-th electron and its ion core, respectively. The last term in
the curly bracket in equation (\ref{eq:hamiltonian}) is the only
difference from the phonon model. We apply the periodic boundary condition,
i.e., $q_{N+1}\equiv q_{1}$. Let the plasmon coordinate and momentum
operators be\begin{equation}
Q_{k}=\frac{1}{\sqrt{N}}\sum_{s}q_{s}e^{-iksa},P_{k}=\frac{1}{\sqrt{N}}\sum_{s}p_{s}e^{iksa}\end{equation}
where $k=2\pi n/Na$, $n=0,1,\cdots,N-1$. (We use notations $k,n,N$
that are used in other part of this paper, since there is no danger
of confusion.) Following the standard procedure, we have

\begin{equation}
H=\sum_{k}\left\{ \frac{P_{k}P_{-k}}{2m_{0}}+\frac{m_{0}\omega_{k}^{2}}{2}Q_{k}Q_{-k}\right\} \label{eq:hamiltonian2}\end{equation}
where \begin{equation}
\omega_{k}^{2}=\frac{1}{m_{0}}\left[2C\left(1-\cos ka\right)+C'\right]\end{equation}
Though the model is crude, this reproduces the usual plasmon dispersion
relation $\omega_{k}\propto k^{2}+constant$ as $k\rightarrow0$.
The standard method of quantization for phonons carries over to the
case of plasmons, for which the creation operator is\begin{equation}
c_{k}^{\dagger}=\frac{1}{\sqrt{2\hbar}}\left[\sqrt{m_{0}\omega_{k}}Q_{-k}-\frac{i}{\sqrt{m_{0}\omega_{k}}}P_{k}\right]\end{equation}
Now, consider the weak coupling limit $C\rightarrow0$. The creation
operator for the individual 'bare' harmonic oscillator at the $s$-th
site is

\begin{equation}
a_{s}^{\dagger}=\frac{1}{\sqrt{2\hbar}}\left[\sqrt{m_{0}\omega_{b}}q_{s}-\frac{i}{\sqrt{m_{0}\omega_{b}}}p_{s}\right]\end{equation}
where the bare oscillation frequency $\omega_{b}$ is given by $\omega_{b}^{2}=C'/m_{0}$.
When $C\left(ka\right)^{2}\ll C'$, or in other words at the large
wavelength limit, we have $\omega_{b}\cong\omega_{k}$. In this case,
combining some of the above relations, we obtain\begin{equation}
c_{k}^{\dagger}=\frac{1}{\sqrt{N}}\sum_{s}\frac{1}{\sqrt{2\hbar}}\left[\sqrt{m_{0}\omega_{k}}q_{s}-\frac{i}{\sqrt{m_{0}\omega_{k}}}p_{s}\right]e^{iksa}\cong\frac{1}{\sqrt{N}}\sum_{s}a_{s}^{\dagger}e^{iksa}\label{eq:creation operator relation}\end{equation}
What equation (\ref{eq:creation operator relation}) states is the
following. \emph{An excitation of a plasmon is equivalent to the superposition
of excitations of each scattering atom, with an additional phase factor
that corresponds to the phase of the plasmon wave} at least in the
initial stage of time evolution. Although our model is crude, we will
henceforth assume that the preceding statement is robust so that it
may be applied to real situations with reasonable degree of validity.
From experiments we know that inelastic scattering is associated with
a small scattering angle that corresponds to delocalization and long
wave length plasmons. Thus, unlike the case of inner-shell electron
excitations, when one atom is excited in an inelastic scattering process
generating a plasmon, the state must be superposed with other states,
in each of which another atom, i.e. one of the surrounding atoms,
is excited. 

In order to see how an excitation of a plasmon affects the measurement,
we consider scattering of a plane wave. This suffices because, first,
we can separately consider two terms of the whole wavefunction, containing
respectively the CPB state $|0>_{b}$ and $|1>_{b}$, and later add
the results. Second, all possible incident electron waves, including
the diverging spherical wave, can be expanded as a sum of plane waves.
We now consider all atoms that contribute their valence electron to
form plasmons. For definiteness, let the number of these atoms be
$N$. We label them with an integer $s=1,\cdots,N$ (changing the
meaning of $N$ somewhat) and, for simplicity, we consider only the
ground state $|g>_{s}$ and the first excited state $|e>_{s}$ for
each of them. Let the position of the $s$-th atom be $\boldsymbol{r}_{s}$.
Let the incident plane wave be $|0>$. Since there are many quantum
entities, we will generally omit the tensor product symbol $\otimes$.
The state before inelastic scattering is\begin{equation}
|\varphi_{1}>=|0>|g>_{1}|g>_{2}\cdots|g>_{N}\end{equation}
We confine ourselves to the case of single plasmon generation and
we neglect terms with two or more atoms excited. A few definitions
follow. Let $|p_{s}>$ denote the electron state after interacting
with the $s$-th atom, without exciting the atom's electron. (One
may view this state as a superposition of the transmitted wave and
elastically scattered wave.) We let the state $|q_{s}>$ represent
the electron state after exciting the atom. Then, after inelastic
scattering, the state becomes\[
\sqrt{N}|\varphi_{2}>=\left\{ \left(c|p_{1}>|g>_{1}+d|q_{1}>|e>_{1}\right)|g>_{2}|g>_{3}\cdots|g>_{N}\right\} \]
\[
+\left\{ |g>_{1}\left(c|p_{2}>|g>_{2}+d|q_{2}>|e>_{2}\right)|g>_{3}|g>_{4}\cdots|g>_{N}\right\} \]
\begin{equation}
+\cdots+\left\{ |g>_{1}|g>_{2}\cdots|g>_{N-1}\left(c|p_{N}>|g>_{N}+d|q_{N}>|e>_{N}\right)\right\} \label{eq:many atoms excited}\end{equation}
where $c,d$ are probability amplitudes to reach the states $|p_{s}>,|q_{s}>$.
(For simplicity we assume that $c,d$ are independent of $s$. This
may not be justified, but it will be found later that this will not
affect our conclusion.) For brevity, we define some of the state of
the whole specimen atoms\[
|G>\equiv|g>_{1}|g>_{2}\cdots|g>_{N},|E_{1}>\equiv|e>_{1}|g>_{2}\cdots|g>_{N},\]
\begin{equation}
|E_{2}>\equiv|g>_{1}|e>_{2}|g>_{3}\cdots|g>_{N},\cdots,|E_{N}>\equiv|g>_{1}|g>_{2}\cdots|g>_{N-1}|e>_{N}\end{equation}
and also define an electron state $|P>\equiv|p_{1}>+|p_{2}>+\cdots+|p_{N}>$.
In terms of these, equation (\ref{eq:many atoms excited}) is rewritten
as\begin{equation}
\sqrt{N}|\varphi_{2}>=c|P>|G>+d\left(|q_{1}>|E_{1}>+|q_{2}>|E_{2}>+\cdots+|q_{N}>|E_{N}>\right)\end{equation}
The first term corresponds to elastic scattering, with the specimen
left without excitation. Now, according to the argument in the last
paragraph, the state of a plasmon with wave vector $\boldsymbol{k}$
is expressed as \begin{equation}
|\boldsymbol{k}>=\frac{1}{\sqrt{N}}\left(e^{i\boldsymbol{k}\cdot\boldsymbol{r}_{1}}|E_{1}>+e^{i\boldsymbol{k}\cdot\boldsymbol{r}_{2}}|E_{2}>+\cdots+e^{i\boldsymbol{k}\cdot\boldsymbol{r}_{N}}|E_{N}>\right)\end{equation}
Although the atoms are not regularly placed, by analogy with the Fourier
transform we write\begin{equation}
|E_{s}>\cong\frac{1}{\sqrt{N}}\sum_{\boldsymbol{k}}e^{-i\boldsymbol{k}\cdot\boldsymbol{r}_{s}}|\boldsymbol{k}>\end{equation}
Hence we have \begin{equation}
\sqrt{N}|\varphi_{2}>\cong c|P>|G>+\frac{d}{\sqrt{N}}\sum_{\boldsymbol{k}}\left\{ \sum_{s=1}^{N}e^{-i\boldsymbol{k}\cdot\boldsymbol{r}_{s}}|q_{s}>\right\} |\boldsymbol{k}>\end{equation}
Therefore, if we observe the plasmon in the state $|\boldsymbol{k}>$
(though this 'observation' only models the decoherence process and
we do not obtain the outcome of such an 'observation' in practice),
the electron wave is left in the state\begin{equation}
|\psi>\cong\frac{1}{\sqrt{N}}\sum_{s=1}^{N}e^{-i\boldsymbol{k}\cdot\boldsymbol{r}_{s}}|q_{s}>\label{eq:scattered by a plasmon}\end{equation}
This is a superposition of electron waves inelastically scattered
at each atoms, with a phase factor $e^{-i\boldsymbol{k}\cdot\boldsymbol{r}_{s}}$
depending on the position on the specimen. If there are sufficiently
many atoms, more-or-less uniformly distributed, then equation (\ref{eq:scattered by a plasmon})
describes nothing but a tilted plane wave, whose wave vector is slightly
changed by emitting a plasmon. This agrees with the standard picture
of momentum conservation.

At this point we make a few remarks. First, if the amplitude $d$
depends on $s$ (for example, because $d$ depends on atomic number
\emph{Z}), then we may divide the atoms into classes with the same
$d$ values. As long as the number of atoms in each class is sufficiently
large to retain the above Fourier-transform-like property, the conclusion
stands. Conversely, one may simply neglect a class that comprises
too few of atoms. Second, one may argue that 'observing' the plasmon
with respect to the measurement basis $\left\{ |\boldsymbol{k}>\right\} $
is rather arbitrary and the use of other basis should be analyzed
as well. However, there are no interactions between the electron and
the specimen after scattering, and statistical measurement outcomes
regarding the scattered electron should not depend on how the plasmon
state is 'measured'. (Of course, being entangled to each other, the
'measurement outcomes' on the electron and the plasmon are \emph{correlated}.)
In addition, we will find the use of the basis $\left\{ |\boldsymbol{k}>\right\} $
highly convenient. Third, as far as there are many atoms in the regions
$S_{0}$ and $S_{1}$, all these atoms contribute to the scattered
wave amplitude and there will not be much variation of the wave amplitude
within the respective regions. Hence, unlike the case of inner-shell
electron excitations, we do not expect to have much amplitude error.
Fourth, we may say that the effect of having plasmon scattering is
equivalent to have a thin 'wedge prism' right next to the specimen.
Every time inelastic scattering occurs, the rotation angle $0\leq\chi<2\pi$
of the 'wedge prism' around the optical axis is completely random.
The 'angle between two planes of the wedge' is also random but generally
obeys the formula (\ref{eq:bethe formula}) that specifies the distribution
of inelastic scattering angles. When viewed in this way, it should
be clear that no matter what incident beam we use, be it focused Gaussian
or diverging Gaussian, the effect of plasmon generation is to tilt
the exit wave from the specimen by an angle proportional to the size
of the plasmon wave vector.

The phase error due to plasmon excitations is more problematic than
the amplitude error. The mechanism for the error is that the scattered
electron wave is tilted by the aforementioned 'wedge prism' and hence
the phase compensation step (See Sec. \ref{sub:A-brief-review}) cannot
be done accurately. Let us first roughly estimate the extent of such
an error. The formula (\ref{eq:bethe formula}) is in a sense misleading
because the average scattering angle is much larger than the 'characteristic
angle' $\theta_{E}=41\mu\mathrm{rad}$. In fact, the mean scattering
angle does not even exist for the Lorentzian distribution $\left(\theta^{2}+\theta_{E}^{2}\right)^{-1}$
unless we introduce a cutoff angle $\theta_{c}$, above which there
is no scattering. The angle for the Bethe-ridge $\sqrt{2\theta_{E}}$
is usually taken as a cutoff, which in our case is $\theta_{c}=9.1\mathrm{mrad}$.
Let us compute the fraction $f$ of inelastically scattered electrons
with the scattering angle less than $\theta_{A}$. This is given by
\begin{equation}
f=\int_{0}^{\theta_{A}}\frac{2\pi\sin\theta d\theta}{\theta^{2}+\theta_{E}^{2}}/\int_{0}^{\theta_{c}}\frac{2\pi\sin\theta d\theta}{\theta^{2}+\theta_{E}^{2}}\end{equation}
Approximating $\sin\theta\cong\theta$, we get\begin{equation}
f=\frac{\ln\left[1+\left(\theta_{A}/\theta_{E}\right)^{2}\right]}{\ln\left[1+\left(\theta_{c}/\theta_{E}\right)^{2}\right]}\label{eq:angular fraction}\end{equation}
From this, one may compute, for example, that half of all inelastically
scattered electrons have the scattering angle larger than $\tilde{\theta}=0.61\mathrm{mrad}$.
In other words, this is the median scattering angle and it is much
larger than $\theta_{E}$. A similar calculation yields the mean scattering
angle $\overline{\theta}=1.8\mathrm{mrad}$. These numbers suggests
that the probability distribution is quite unusual due to the long
tail. 

We estimate the phase error due to failure in performing the phase
compensation step. First, consider the case of focused incident beam
(Sec. \ref{sub:Focussed-incident-beam}). As in Appendix A, we take
$z$-axis to be the optical axis. Let the regions $S_{0}$ and $S_{1}$
be separated by a distance $d_{01}=1.0\mathrm{nm}$ along $x$-axis.
Let the pixel, in which the scattered electron is detected, be $ $A.
Let the distance between the point A and the central points of the
specimen regions $S_{0}$ and $S_{1}$ be $L_{0}$ and $L_{1}$ respectively.
Then, the optical path length difference is $\Delta L=\left|L_{1}-L_{0}\right|$.
The detection pixel is placed on the $xz$ plane without loss of generality.
The scattering angle is $\theta$. Simple geometrical consideration
then yields $\Delta L\cong d_{01}\sin\theta\cong d_{01}\theta$. Now,
suppose that a plasmon is generated upon scattering, to the direction
of $x$-axis. The path length difference is modified because of the
presence of aforementioned 'wedge prism', to be $\Delta L\cong d_{01}\sin\left(\theta+\delta\theta\right)\cong d_{01}\left(\theta+\delta\theta\right)$
where $\delta\theta$ is the angle of the 'wedge prism', which actually
is the inelastic scattering angle. The error in estimating the optical
path length difference in the phase compensation step would therefore
be $d_{01}\overline{\theta}=1.8\mathrm{pm}$, if we take mean angle
error $\overline{\theta}$ for plasmon scatterings as the angle of
the 'wedge prism' $\delta\theta$. The error, in terms of phase angle,
is unacceptable $kd_{01}\overline{\theta}=5.7\mathrm{rad}$, where
$k=3.2\mathrm{pm^{-1}}$ is the wave number of $300\mathrm{keV}$
electrons. In addition to the error due to the 'speckle pattern' problem
mentioned in Sec. \ref{sub:Focussed-incident-beam}, this constitutes
another reason why the focused-incident-beam configuration is not
desirable.

Let us turn to the case of diverging incident beam, described in Sec.
\ref{sub:Diverging-incident-beam}. Figure 3 shows that the Gaussian
electron beam focused at either $52.5\mathrm{nm}$ or $22.5\mathrm{nm}$
above the specimen central plane CP. The focal points corresponding
to the regions $S_{0}$ and $S_{1}$ are labeled with the same symbol.
The distance between these focal points is denoted as $d'_{01}=30\mathrm{nm}$.
Let $\theta$ be the scattering angle and let $\delta\theta$ be the
deflection angle due to plasmon generation. The plasmon scattering
occurs at CP in the figure, but exactly where the scattering occurs
does not matter, as the reader will see shortly. We assume that the
plasmon momentum vector is parallel to $x$-axis, which is the worst
case. The four 'optical rays' A, B, C and D go to the same direction
under CP. The rays A and B represents optical paths taken by electrons
that are not inelastically scattered, whereas the rays C and D shows
optical paths associated with plasmon generation. In computing the
optical path length difference, we always make the approximation $\sin\theta\cong\theta$
for any angle. Then, computation of the optical path length difference
is a matter of simple exercise in geometry. The optical path length
difference in the absence of inelastic scattering, between rays A
and B is $d_{01}\theta^{2}/2$. On the other hand, the three thick
portions of rays in Fig. 3 contributes the path length difference
between rays C and D, which turns out to be $d_{01}\left(\theta^{2}-\delta\theta^{2}\right)/2$.
Thus, the change of the optical path length difference due to the
plasmon scattering is $d_{01}\delta\theta^{2}/2$. The magnitudes
of $\delta\theta$ is about the median inelastic scattering angle
$\tilde{\theta}$ or the mean $\bar{\theta}$, depending on how one
defines the 'typical' scattering angle. The error in terms of phase,
then, is $kd_{01}\tilde{\theta}^{2}/2\cong18\mathrm{mrad}$ or $kd_{01}\bar{\theta}^{2}/2\cong155\mathrm{mrad}$.
The latter is not small and requires further investigation.

\begin{figure}
\begin{center}
\includegraphics[width=0.45\textwidth]{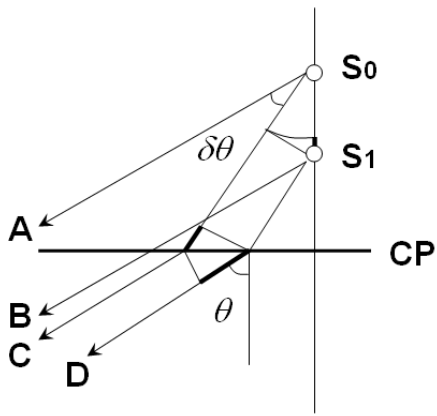}

\caption{A diagram to compute the path length difference associated with a
plasmon scattering in the diverging incident beam case. The angles
$\theta$ and $\delta\theta$ are much exaggerated.}

\end{center}
\end{figure}

In order to resolve the question of how plasmon scattering affects
the measurement, we resort to numerical simulation. We employ $k=36$
electrons in a k-electron process. We estimate the frequency of the
plasmon scattering as follows. At near the end of Sec. \ref{sub:Focussed-incident-beam},
it was found that the probability of elastic scattering for $300\mathrm{keV}$
electrons is $5.2\%$ for a $30\mathrm{nm}$ thick specimen. On the
other hand, generally inelastic scattering is considered to occur
twice as frequent as elastic scattering as mentioned earlier. Hence
we take a value of $p_{inel}=10\%$ as inelastic scattering probability.
We will ignore the failure due to high-angle elastic scattering (probability
$4.5\times10^{-3}$ as computed earlier) and inner-core electron excitations
(probability $8.6\times10^{-4}$), as these are sufficiently rare.
Indeed, when these probabilities are added we have $p_{fail}=0.54\%$.
A simple consideration shows that the k-electron process fails with
a probability $\cong kp_{fail}=19\%$. A $19\%$ waste of electron
dose is acceptable. 

Following Ref. \cite{ent-assisted EM}, we consider cryoelectron microscopy
of the ribosome molecule. However, unlike Ref. \cite{ent-assisted EM}
we consider $300\mathrm{keV}$ electrons. Otherwise the method of
simulation is the same as Ref. \cite{ent-assisted EM}, except that
we now consider inelastic scattering that occurs $10\%$ of times
at random. When it is decided in our simulation program that inelastic
scattering happened, the phase error is computed as follows. First,
a uniform random number $r$ in the interval $\left[0,1\right]$ is
generated. Then we use equation (\ref{eq:angular fraction}) by letting
$f=r$ and compute $\theta_{A}$. Clearly, $\theta_{A}$ obeys the
statistical distribution governing inelastic scattering angle. On
the other hand, suppose that the plasmon wave vector is proportional
to $\cos\chi\boldsymbol{e}_{x}+\sin\chi\boldsymbol{e}_{y}$, where
$\boldsymbol{e}_{x},\boldsymbol{e}_{y}$ are unit vectors parallel
to $x,y$ axes, respectively. Geometrical consideration shows that
the phase error in angle is $\varphi_{err}\cong kd'_{01}\theta_{A}^{2}\cos\chi$,
where $\chi$ is a random number drawn uniformly from the interval
$\left[0,2\pi\right]$. Since this error adds to the phase we want
to measure during a k-electron process, equation (\ref{eq:CPB reading prob})
is now modified as\begin{equation}
P_{\leftarrow}=\frac{1+\sin\left(k\Delta\varphi+\varphi_{tot}\right)}{2},\textrm{ }P_{\rightarrow}=1-P_{\leftarrow}.\label{eq:CPB reading prob2}\end{equation}
where $\varphi_{tot}$ represents the total error accumulated in the
k-electron process. It is the sum of all individual error $\varphi_{err}$
occurred during the k-electron process. More precisely, if $n_{inel}$
inelastic scattering happened in a k-electron process, then there are
$n_{inel}$ independent random variables $\varphi_{err}$, that are
summed up to produce $\varphi_{tot}$.

Figure 4 shows the simulation result. Each image consists of $100\times100$
pixels and the size of the image is $30\mathrm{nm}\times30\mathrm{nm}$.
The size of Gaussian beams to the regions $S_{0},S_{1}$ are respectively
$3.0\mathrm{nm}$ and $0.6\mathrm{nm}$ at the entrance surface of
the specimen. As mentioned in Sec. \ref{sub:Diverging-incident-beam},
there is a significant beam spread in the specimen but we did \emph{not}
take the spread into account in the simulation, which has been performed
as if there were no such spread \cite{Comment 4}. The electron dose
is $400e\mathrm{/nm^{2}}$. Perfect detector efficiency and the CPB
readout efficiency is assumed. The 'raw' image is smoothed with a
Gaussian filter with a standard deviation $0.3\mathrm{nm}$ for better
visibility. A simulated image of entanglement-assisted electron microscopy,
which takes errors due to inelastic scattering into account, is shown
in Fig. 4 (a). For the sake of comparison, a 'Laplacian-filtered'
\cite{Comment 5} phase shift map of the ribosome, from a X-ray study
\cite{Ban ribosome}, is shown in Fig. 4 (b); the hypothetical case
where inelastic scattering is absent, but otherwise identical situation
, is shown in Fig. 4 (c); and an image for the 'conventional' case,
where perfect in-focus phase contrast microscopy is employed with
the same dose, also 'Laplacian-filtered', is shown in Fig. 4 (d).

\begin{figure}
\begin{center}
\includegraphics[width=0.23\textwidth]{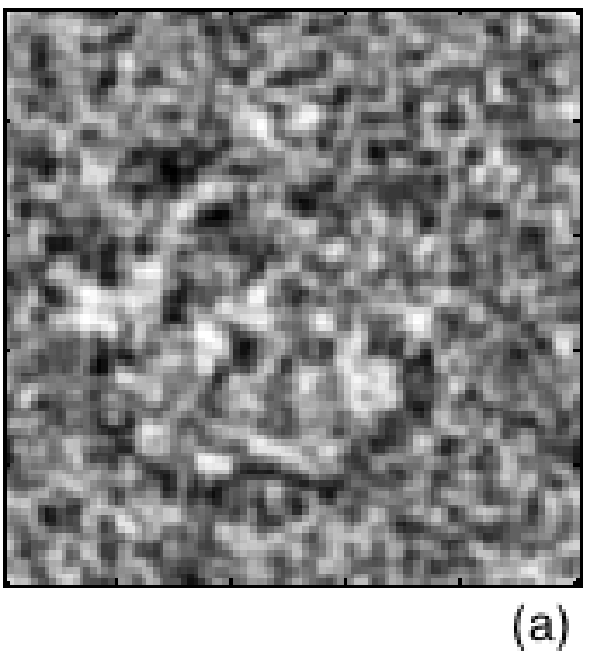}\includegraphics[width=0.23\textwidth]{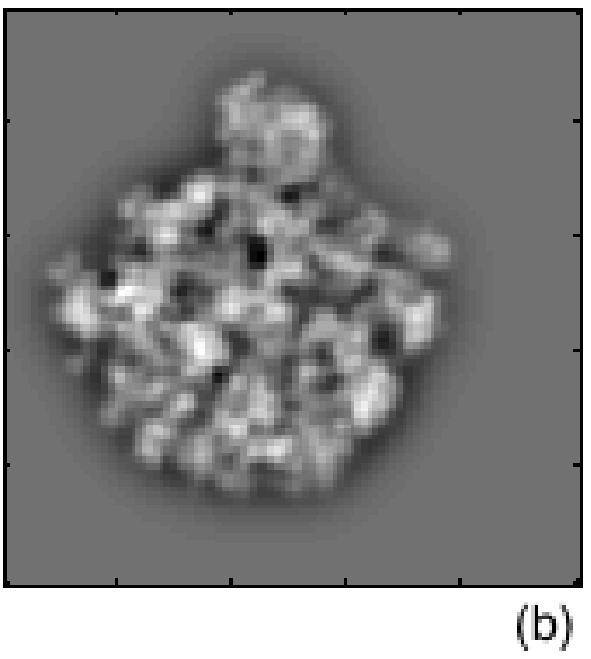}\includegraphics[width=0.23\textwidth]{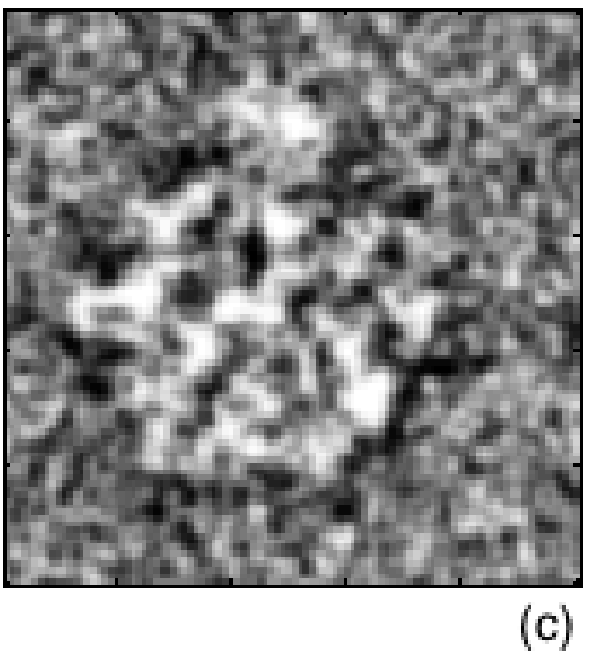}\includegraphics[width=0.23\textwidth]{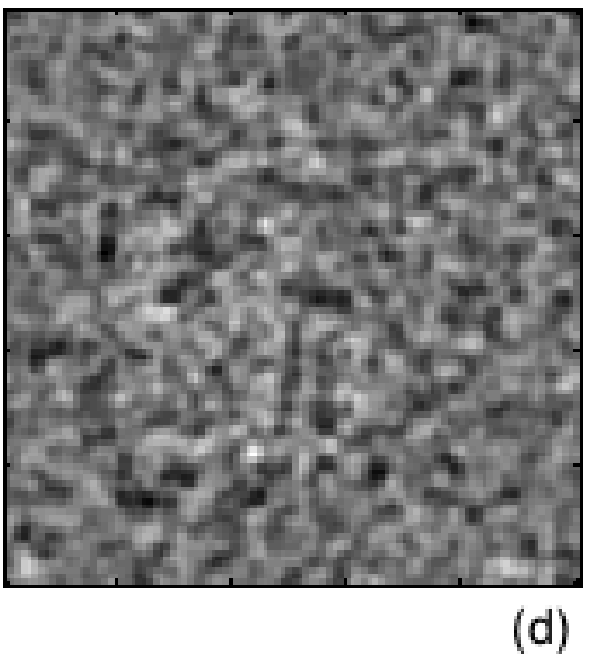}

\caption{Comparison of simulated images. (a) Entanglement assisted electron
microscopy with the effect of inelastic scattering taken into account.
(b) The phase map of the ribosome molecule determined by X-ray crystallography.
(c) A hypothetical case with entanglement-assisted electron microscopy
where there is no inelastic scattering. (d) The 'conventional' case
of perfect in-focus phase contrast microscopy. The size of all images is $(30\mathrm{nm})^2$.}

\end{center}
\end{figure}

\section{Conclusion\label{sec:Conclusion}}

We evaluated measurement errors in entanglement-assisted electron
microscopy, due both to elastic and inelastic scattering processes
at the specimen. Despite remaining subtleties such as processing of
data obtained by the significantly divergent electron beam, overall
we have found no evidence that entanglement-assisted electron microscopy
is not viable for some fundamental physical reasons related to electron
scattering processes. There are several lessons learned along the
way. First, despite the fact that lossy processes are often emphasized
as the sole fundamental limit in the quantum metrology literature,
elastic scattering processes present a limit just as fundamental in
the present case. The reason is that the elastic scattering signal
sometimes contains information that we do not need, because the information
is about too high resolution features of the specimen that cannot
be fully obtained anyway. This leads to an error which cannot be controlled
nor corrected, but it has been computed to be sufficiently small.
Second, besides the inner-shell excitation processes associated with
a sufficiently small cross sections, inelastic scattering processes
involve generation of plasmons. The excitation of a plasmon results
in bending of the probe electron wave, which in turn entails phase
error because the phase compensation step of the entanglement-assisted
electron microscopy cannot be carried out precisely. We found that
the magnitude of such an error is acceptable. Rather remarkably, how
the two probe electron states (that are entangled to the two CPB states)
are geometrically arranged is found to be important in order to reduce
this type of error. 

Many issues are left for future investigations. For example, the most
suitable energy of the primary electron beam, which presumably depends
on the specimen thickness, has not been fully understood, although
the present author feels that higher energy is better, unless knock-on
damage comes in to play a significant role. Furthermore, errors associated
with the electron reflection process at the electron mirror equipped
with the CPB, or errors associated with the CPB control processes,
have largely been unexplored.
\subsection*{Acknowledgments}
I thank Umpei Miyamoto for his help with TEX. This research was supported in part by the JSPS {}``Kakenhi'' Grant
(\#25390083).

\section*{APPENDIX A: EXPRESSIONS OF THE ELASTIC SCATTERING AMPLITUDE\label{sec:APPENDIX-A:-EXPRESSIONS}}

Here we derive equations (\ref{eq:transmitted wave far field}), (\ref{eq:scattered wave 1})
and (\ref{eq:scattered wave 2}) appeared in Sec. \ref{sec:Coherent errors}
and clarify what approximations are involved in the derivation. Let
us first deal with the case of Sec. \ref{sub:Focussed-incident-beam}.
The incident Gaussian wave is a superposition of many plane waves.
For a constituent plane incident wave $e^{i\boldsymbol{k}_{i}\cdot\boldsymbol{r}}$,
after elastic scatterings we have, in the far field \begin{equation}
e^{i\boldsymbol{k}_{i}\cdot\boldsymbol{r}}+\sum_{s}\frac{f_{s}\left(\theta_{\boldsymbol{k}_{i},\boldsymbol{k}_{f}}\right)}{\left|\boldsymbol{r}-\boldsymbol{r}_{s}\right|}e^{i\boldsymbol{k}_{i}\cdot\boldsymbol{r}_{s}}e^{ik\left|\boldsymbol{r}-\boldsymbol{r}_{s}\right|}\label{eq:plane wave scattering-1}\end{equation}
where $\boldsymbol{k}_{f}$ is the 'wave number vector after the scattering',
whose primary purpose is to point at the point P, where we wish to
compute the value of the wavefunction, and $\theta_{\boldsymbol{k}_{i},\boldsymbol{k}_{f}}$
is the angle between the vectors $\boldsymbol{k}_{i}$ and $\boldsymbol{k}_{f}$.
We are dealing with elastic scattering and $\left|\boldsymbol{k}_{i}\right|=\left|\boldsymbol{k}_{f}\right|=k$,
at least to a good approximation. We ignore the imaginary part of
$f_{s}$, as is done in the Born approximation though this cannot
be entirely correct because of the Optical Theorem. As usual, when
$\left|\boldsymbol{r}-\boldsymbol{r}_{s}\right|$ appearing in the
denominator in eq. (\ref{eq:plane wave scattering-1}) is much greater
than all relevant $\left|\boldsymbol{r}_{s}\right|$, $\left|\boldsymbol{r}-\boldsymbol{r}_{s}\right|$
may be approximated to be $r=\left|\boldsymbol{r}\right|$. This is
a representative distance between the whole specimen and the point
P. Let the $x,y,z$ component of the vector $\boldsymbol{k}_{i}$
be respectively $k_{x},k_{y},k_{z}$, which we write $\boldsymbol{k}_{i}=\left(k_{x},k_{y},k_{z}\right)$.
We want the incident beam to be Gaussian with the waist size $w_{0}$
and the above plane wave solutions are superposed as\begin{equation}
\pi w_{0}^{2}\int\frac{dk_{x}}{2\pi}\int\frac{dk_{y}}{2\pi}e^{-w_{0}^{2}\left(k_{x}^{2}+k_{y}^{2}\right)/4}\left[e^{i\boldsymbol{k}_{i}\cdot\boldsymbol{r}}+\frac{1}{r}\sum_{s}f_{s}\left(\theta_{\boldsymbol{k}_{i},\boldsymbol{k}_{f}}\right)e^{i\boldsymbol{k}_{i}\cdot\boldsymbol{r}_{s}}e^{ik\left|\boldsymbol{r}-\boldsymbol{r}_{s}\right|}\right]\label{eq:Gaussian wave scattering-1}\end{equation}
The factor $\pi w_{0}^{2}$ is introduced solely to simplify some
of later expressions and keep them dimensionless. The diffraction-limited
waist of the Gaussian beam is on the specimen. At $z=0$, the incident
beam part is\begin{equation}
\pi w_{0}^{2}\int_{-\infty}^{\infty}\frac{dk_{x}}{2\pi}\int_{-\infty}^{\infty}\frac{dk_{y}}{2\pi}e^{-w_{0}^{2}\left(k_{x}^{2}+k_{y}^{2}\right)/4}e^{i\left(k_{x}x+k_{y}y\right)}=e^{-\left(x^{2}+y^{2}\right)/w_{0}^{2}}\label{eq:incident beam-1}\end{equation}
Note that the $z$ component of the $\boldsymbol{k}_{i}$-vector is
$k_{z}=\sqrt{k^{2}-k_{x}^{2}-k_{y}^{2}}$ in above equations and the
limits of integrations in equation (\ref{eq:incident beam-1}) are
approximated as positive and negative infinities. Far from the specimen,
the transmitted Gaussian wave has the known form \begin{equation}
\frac{-ikw_{0}^{2}}{2z}\exp\left[-\frac{k^{2}w_{0}^{2}}{4z^{2}}\left(x^{2}+y^{2}\right)+ik\left(z+\frac{x^{2}+y^{2}}{2z}\right)\right]\cong\frac{-ikw_{0}^{2}}{2z}\exp\left[-\frac{k^{2}w_{0}^{2}}{4z^{2}}\left(x^{2}+y^{2}\right)\right]e^{ikr}\label{eq:transmitted wave far field-1}\end{equation}
This, when $y=0$, is equation (\ref{eq:transmitted wave far field})
of the main text.

Next, we consider the scattered part of the wave amplitude\begin{equation}
\frac{\pi w_{0}^{2}}{r}\sum_{s}\left[\int_{-\infty}^{\infty}\int_{-\infty}^{\infty}\frac{dk_{x}}{2\pi}\frac{dk_{y}}{2\pi}e^{-w_{0}^{2}\left(k_{x}^{2}+k_{y}^{2}\right)/4}f_{s}\left(\theta_{\boldsymbol{k}_{i},\boldsymbol{k}_{f}}\right)e^{i\boldsymbol{k}_{i}\cdot\boldsymbol{r}_{s}}e^{ik\left|\boldsymbol{r}-\boldsymbol{r}_{s}\right|}\right]\label{eq:scattered wave-1}\end{equation}
Note that, if the phase factor $e^{i\boldsymbol{k}_{i}\cdot\boldsymbol{r}_{s}}e^{ik\left|\boldsymbol{r}-\boldsymbol{r}_{s}\right|}$
were identically $1$, which it is not, the integration in the square
brackets in equation (\ref{eq:scattered wave-1}) would represent,
up to a constant multiplicative factor, something akin to, but not
identical with, Gaussian smoothing of $f_{s}\left(\theta_{\boldsymbol{k}_{i},\boldsymbol{k}_{f}}\right)$. 

To evaluate the phase factor $e^{i\boldsymbol{k}_{i}\cdot\boldsymbol{r}_{s}}e^{ik\left|\boldsymbol{r}-\boldsymbol{r}_{s}\right|}$,
we write it as $e^{i\boldsymbol{k}_{i}\cdot\boldsymbol{r}_{s}}e^{ik\left(\left|\boldsymbol{r}-\boldsymbol{r}_{s}\right|-r\right)}e^{ikr}$.
Let us first consider the exit wave, and focus on a single pixel of
the detector. Since the pixel is on a diffraction plane, $\boldsymbol{k}_{f}$
is independent of $s$. In other words, it is independent of the place
of the atom under consideration. Without loss of generality, we assume
that the pixel under consideration is on the $xz$ plane. Let the
angle between $z$-axis and $\boldsymbol{k}_{f}$ be $\theta$. Then
the detector pixel is located at $\boldsymbol{r}=\left(r_{x},0,r_{z}\right)=r\left(\sin\theta,0,\cos\theta\right)$,
where $r$ is much greater than any $r_{s}$. Writing $\boldsymbol{r}_{s}=\left(x_{s},y_{s},z_{s}\right)$,
we have \begin{equation}
\left|\boldsymbol{r}-\boldsymbol{r}_{s}\right|-r=\sqrt{r^{2}+r_{s}^{2}-2\left(r_{x}x_{s}+r_{z}z_{s}\right)}-r\cong-\left(x_{s}\sin\vartheta+z_{s}\cos\vartheta\right)\end{equation}
Let us now turn to the incident wave and write $\boldsymbol{k}_{i}=k\left(\sin\delta_{x},\sin\delta_{y},\cos\delta\right)$.
It is clear that the angle between $\boldsymbol{k}_{i}$ and $z$-axis
is $\delta$, $\sin^{2}\delta_{x}+\sin^{2}\delta_{y}=\sin^{2}\delta$,
and hence $\delta_{x}^{2}+\delta_{y}^{2}\cong\delta^{2}$ when these
angles are small. Putting them together, the phase factor $e^{i\boldsymbol{k}_{i}\cdot\boldsymbol{r}_{s}}e^{ik\left(\left|\boldsymbol{r}-\boldsymbol{r}_{s}\right|-r\right)}e^{ikr}$
equals\[
\exp\left[ik\left(x_{s}\left(\sin\delta_{x}-\sin\theta\right)+y_{s}\sin\delta_{y}+z_{s}\left(\cos\delta-\cos\theta\right)\right)\right]e^{ikr}\]

\begin{equation}
\cong\exp\left[ik\left(x_{s}\delta_{x}+y_{s}\delta_{y}-z_{s}\frac{\delta^{2}}{2}\right)\right]\exp\left[ik\left(-x_{s}\theta+z_{s}\frac{\theta^{2}}{2}\right)\right]e^{ikr}\label{eq:phase factor-1}\end{equation}
This makes equation (\ref{eq:scattered wave-1})\begin{equation}
\frac{\pi w_{0}^{2}e^{ikr}}{r}\sum_{s}e^{ik\left(-x_{s}\theta+z_{s}\frac{\theta^{2}}{2}\right)}\left[\int_{-\infty}^{\infty}\int_{-\infty}^{\infty}\frac{dk_{x}}{2\pi}\frac{dk_{y}}{2\pi}e^{-w_{0}^{2}\left(k_{x}^{2}+k_{y}^{2}\right)/4}f_{s}\left(\theta_{\boldsymbol{k}_{i},\boldsymbol{k}_{f}}\right)e^{ik\left(x_{s}\delta_{x}+y_{s}\delta_{y}-z_{s}\frac{\delta^{2}}{2}\right)}\right]\label{eq:scattered wave 2-1}\end{equation}

In order to further approximate, we check some numbers here. Because
of the waist size $w_{0}\sim0.5\mathrm{nm}$ (since we want a $\sim1\mathrm{nm}$
\emph{diameter}), $x_{s}$ and $y_{s}$ are at most $\sim1\mathrm{nm}$,
while $z_{s}$ is up to the specimen thickness $\pm15\mathrm{nm}$.
From the theory of Gaussian beams we know that the wave convergence
half-angle is $\theta_{G}=2/kw_{0}$, where the wave number $k$ is
$3.2\times10^{3}\mathrm{nm}^{-1}$ for $300\mathrm{keV}$ electrons.
This makes $\theta_{G}\cong1.3\mathrm{mrad}$ (and the 'focal depth',
or the Rayleigh range $z_{R}$, is $kw_{0}^{2}/2=w_{0}/\theta_{G}=0.4\mu\mathrm{m}$).
Noting that $\delta_{x},\delta_{y},\delta$ is at most of the order
$\theta_{G}$, the phase factor in the square bracket in equation
(\ref{eq:scattered wave 2-1}) may be approximated as $e^{ik\left(x_{s}\delta_{x}+y_{s}\delta_{y}\right)}\cong e^{i\left(k_{x}x_{s}+k_{y}y_{s}\right)}$.
On the other hand, the meaningful magnitude of $\theta$ should roughly
be given by the {}``characteristic angle of elastic scattering''
$\theta_{0}=1/kr_{0}$, where $r_{0}$ is the screening radius of
the atom and is given by $r_{0}=a_{0}Z^{-1/3}$, where $a_{0}$ and
$Z$ are the Bohr radius and the atomic number, respectively \cite{Egerton textbook}.
For most relevant elements (See Appendix B) $Z^{-1/3}\cong0.5$ and
we have $r_{0}\cong26\mathrm{pm}$ and $\theta_{0}\cong10\mathrm{mrad}$
and in the case of hydrogen this is even smaller. These numbers allow
us to further approximate equation (\ref{eq:scattered wave 2-1})
to yield\begin{equation}
\frac{\pi w_{0}^{2}e^{ikr}}{r}\sum_{s}e^{-ikx_{s}\theta}\left[\int_{-\infty}^{\infty}\int_{-\infty}^{\infty}\frac{dk_{x}}{2\pi}\frac{dk_{y}}{2\pi}e^{-w_{0}^{2}\left(k_{x}^{2}+k_{y}^{2}\right)/4}f_{s}\left(\theta_{\boldsymbol{k}_{i},\boldsymbol{k}_{f}}\right)e^{i\left(k_{x}x_{s}+k_{y}y_{s}\right)}\right]\label{eq:scattered wave 3-1}\end{equation}
This approximation should be understood as, for example, keeping only
the phase shift $e^{ikx_{s}\theta}$ associated with atoms with large
$x_{s}$ values. (We have the phase factor $e^{ikx_{s}\theta}$ for
atoms with smaller $x_{s}$ values also, but that may be less significant
than the neglected $e^{ikz_{s}\theta^{2}/2}$.) Furthermore, since
the presence of the factor $e^{-w_{0}^{2}\left(k_{x}^{2}+k_{y}^{2}\right)/4}$
implies that the integration, in terms of the angle $\theta_{\boldsymbol{k}_{i},\boldsymbol{k}_{f}}$,
is over a narrow range of the order $\theta_{G}\cong1.3\mathrm{mrad}$.
This is fairly smaller than the 'spread' of the function $f_{s}\left(\theta_{\boldsymbol{k}_{i},\boldsymbol{k}_{f}}\right)$,
that is $\cong\theta_{0}$, when plotted with respect to $\theta_{\boldsymbol{k}_{i},\boldsymbol{k}_{f}}$.
These facts allow us to approximate equation (\ref{eq:scattered wave 3-1})
to obtain\begin{equation}
\frac{\pi w_{0}^{2}e^{ikr}}{r}\sum_{s}e^{-ikx_{s}\theta}f_{s}\left(\theta\right)\left[\int_{-\infty}^{\infty}\int_{-\infty}^{\infty}\frac{dk_{x}}{2\pi}\frac{dk_{y}}{2\pi}e^{-w_{0}^{2}\left(k_{x}^{2}+k_{y}^{2}\right)/4}e^{i\left(k_{x}x_{s}+k_{y}y_{s}\right)}\right]\label{eq:scattered wave 4-1}\end{equation}
The integral is the two-dimensional Fourier transform of a Gaussian
function and we get \begin{equation}
\frac{e^{ikr}}{r}\sum_{s}e^{-\frac{x_{s}^{2}+y_{s}^{2}}{w_{0}^{2}}}f_{s}\left(\theta\right)e^{-ikx_{s}\theta}\label{eq:scattered wave 5-1}\end{equation}
This is equation (\ref{eq:scattered wave 1}) of the main text.

We now turn to the second case described in Sec. \ref{sub:Diverging-incident-beam}.
The above discussion up to equation (\ref{eq:scattered wave 2-1})
holds also here. We keep the Gaussian beam waist at $z=0$, but the
atoms in the specimen are no longer at $z\cong0$ but at near $z=\Delta z$
that is $52.5\mathrm{nm}$ or $22.5\mathrm{nm}$, depending on whether
the region under consideration is $S_{0}$ or $S_{1}$. The $z$ coordinate
of the $s$-th atom is now $z_{s}=\Delta z+\hat{z}_{s}$, where $\hat{z}_{s}$
is at most $\pm15\mathrm{nm}$. Equation (\ref{eq:scattered wave 2-1})
is then rewritten as\[
\frac{\pi w_{0}'^{2}e^{ikr}}{r}e^{ik\Delta z\frac{\theta^{2}}{2}}\sum_{s}e^{ik\left(-x_{s}\theta+\hat{z}_{s}\frac{\theta^{2}}{2}\right)}\]
\begin{equation}
\times\left[\int_{-\infty}^{\infty}\int_{-\infty}^{\infty}\frac{dk_{x}}{2\pi}\frac{dk_{y}}{2\pi}e^{-\left(\frac{w_{0}'^{2}}{4}+i\frac{\Delta z}{2k}\right)\left(k_{x}^{2}+k_{y}^{2}\right)}f_{s}\left(\theta_{\boldsymbol{k}_{i},\boldsymbol{k}_{f}}\right)e^{i\left(k_{x}x_{s}+k_{y}y_{s}-\hat{z}_{s}\frac{k_{x}^{2}+k_{y}^{2}}{2k}\right)}\right]\label{eq:scattered wave 6-1}\end{equation}
Here, we need to consider the range of $\theta$ up to $\theta_{max}=50\mathrm{mrad}$,
which is the sum of the incident beam divergence $\theta_{G}'$ and
the characteristic angle $\theta_{0}$. The coordinates $x_{s},y_{s}$
are at most $2.7\mathrm{nm}$ and $k_{x}/k,k_{y}/k$ are at most $\sim\theta_{G}'$.
Hence equation (\ref{eq:scattered wave 6-1}) is approximated as\[
\frac{\pi w_{0}'^{2}e^{ikr}}{r}e^{ik\Delta z\frac{\theta^{2}}{2}}\sum_{s}e^{-ikx_{s}\theta}\left[\int_{-\infty}^{\infty}\int_{-\infty}^{\infty}\frac{dk_{x}}{2\pi}\frac{dk_{y}}{2\pi}e^{-\left(\frac{w_{0}'^{2}}{4}+i\frac{\Delta z}{2k}\right)\left(k_{x}^{2}+k_{y}^{2}\right)}f_{s}\left(\theta_{\boldsymbol{k}_{i},\boldsymbol{k}_{f}}\right)e^{i\left(k_{x}x_{s}+k_{y}y_{s}\right)}\right]\]
\[
=\frac{\pi w_{0}'^{2}e^{ikr}}{r}e^{ik\Delta z\frac{\theta^{2}}{2}}\sum_{s}e^{-ikx_{s}\theta}e^{i\frac{k}{2\Delta z}\left(x_{s}^{2}+y_{s}^{2}\right)}\]
\begin{equation}
\times\left[\int_{-\infty}^{\infty}\int_{-\infty}^{\infty}\frac{dk_{x}}{2\pi}\frac{dk_{y}}{2\pi}e^{-\frac{w_{0}'^{2}}{4}\left(k_{x}^{2}+k_{y}^{2}\right)}e^{-i\frac{\Delta z}{2k}\left[\left(k_{x}-\frac{kx_{s}}{\Delta z}\right)^{2}+\left(k_{y}-\frac{ky_{s}}{\Delta z}\right)^{2}\right]}f_{s}\left(\theta_{\boldsymbol{k}_{i},\boldsymbol{k}_{f}}\right)\right]\label{eq:scattered wave 7-1}\end{equation}
though the approximation is not good. Consider the integral in the
square bracket. The phase factor $e^{-i\frac{\Delta z}{2k}\left[\left(k_{x}-\frac{kx_{s}}{\Delta z}\right)^{2}+\left(k_{y}-\frac{ky_{s}}{\Delta z}\right)^{2}\right]}$
has the associated phase angle up to $\sim k\Delta z\left(\theta'_{G}\right)^{2}$.
In our case $k\Delta z\cong10^{5}$ and $k\Delta z\left(\theta'_{G}\right)^{2}\cong160$,
thus one may say that the phase angle rotates fairly rapidly. Hence
the value of $ $$f_{s}\left(\theta_{\boldsymbol{k}_{i},\boldsymbol{k}_{f}}\right)$
matters only at around $k_{x}\Delta z\cong kx_{s},k_{y}\Delta z\cong ky_{s}$,
where no first-order phase oscillation occurs. Since $\boldsymbol{k}_{i}=k\left(x_{s}/\Delta z,y_{s}/\Delta z,1\right)$
and $\boldsymbol{k}_{f}=k\left(\sin\theta,0,\cos\theta\right)$, we
have $\theta_{\boldsymbol{k}_{i},\boldsymbol{k}_{f}}\cong\theta-x_{s}/\Delta z$
to the first order in $x_{s},y_{s},\theta$. Hence, we further simplify
equation (\ref{eq:scattered wave 7-1}) as\[
=\frac{\pi w_{0}'^{2}e^{ikr}}{r}e^{ik\Delta z\frac{\theta^{2}}{2}}\sum_{s}e^{-ikx_{s}\theta}e^{i\frac{k}{2\Delta z}\left(x_{s}^{2}+y_{s}^{2}\right)}f_{s}\left(\theta-\frac{x_{s}}{\Delta z}\right)\]
\begin{equation}
\times\left[\int_{-\infty}^{\infty}\int_{-\infty}^{\infty}\frac{dk_{x}}{2\pi}\frac{dk_{y}}{2\pi}e^{-\frac{w_{0}'^{2}}{4}\left(k_{x}^{2}+k_{y}^{2}\right)}e^{-i\frac{\Delta z}{2k}\left[\left(k_{x}-\frac{kx_{s}}{\Delta z}\right)^{2}+\left(k_{y}-\frac{ky_{s}}{\Delta z}\right)^{2}\right]}\right]\label{eq:scattered wave 8-1}\end{equation}
We must note, however, that this approximation is only barely justifiable
because the phase factor $e^{-i\frac{\Delta z}{2k}\left[\left(k_{x}-\frac{kx_{s}}{\Delta z}\right)^{2}+\left(k_{y}-\frac{ky_{s}}{\Delta z}\right)^{2}\right]}$
is fairly constant in an angular range $\Delta\theta\cong\sqrt{\lambda/\Delta z}$,
where $\lambda=2.0\mathrm{pm}$ is the wavelength of $300\mathrm{keV}$
electrons. Taking $\Delta z=22.5\mathrm{nm}$, we have $\Delta\theta\cong10\mathrm{mrad}$,
which actually is comparable to the characteristic angle $\theta_{0}$
associated with the scattering amplitude $f_{s}$. Due to the lack
of simple and good analytical alternatives we proceed. The integral
is a convolution of two functions that equals\begin{equation}
\frac{-ik}{\pi\left(2\Delta z-iw_{0}'^{2}k\right)}e^{-\frac{k^{2}w_{0}'^{2}}{2\Delta z\left(2\Delta z-iw_{0}'^{2}k\right)}\left(x_{s}^{2}+y_{s}^{2}\right)}\label{eq:convolution-1}\end{equation}
It may be verified that $\Delta z\gg kw_{0}'^{2}/2$ and we define
$\varepsilon\equiv kw_{0}'^{2}/2\Delta z\sim0.02$. (The parameter
$\varepsilon$ has a physical meaning of the ratio between the waist
size of the incident beam $w_{0}'$ and the incident beam 'radius'
at the distance $\Delta z$ from the waist. Alternatively, $\varepsilon=z_{R}/\Delta z$,
where $z_{R}$ is the Rayleigh range.) Hence, equation (\ref{eq:scattered wave 8-1})
approximately equals\begin{equation}
\frac{-i\varepsilon e^{ikr}}{r}\sum_{s}e^{i\frac{k}{2\Delta z}\left[\left(x_{s}-\Delta z\theta\right)^{2}+y_{s}^{2}\right]}e^{-\varepsilon\frac{k}{2\Delta z}\left(x_{s}^{2}+y_{s}^{2}\right)}f_{s}\left(\theta-\frac{x_{s}}{\Delta z}\right)\label{eq:scattered wave 9-1}\end{equation}
This is equation (\ref{eq:scattered wave 2}) in the main text.

\section*{APPENDIX B: ATOMIC NUMBER DENSITIES IN A TYPICAL BIOLOGICAL SPECIMEN}

We consider a frozen, hydrated speimen that is typical in biological
cryoelectron microscopy and compute the associated atomic number densities
for the elements H, C, N, O, and S. The water content may vary considerably
from specimen to specimen. However, for simplicity we take values
of 76\% water and 24\% {}``dry weight'' as typical composition,
which has been reported as the composition of {}``sediments obtained
by predetermined dilution of pellets harvested by centrifugation''
of yeast suspension \cite{water content}. The component which contributes
to the dry weight may in general contain things other than protein.
However, again for simplicity, we assume that the component consists
solely of protein. Then, given the known density of low density amorphous
(LDA) ice 0.94 g/cc \cite{amorphous ice density}, the standard value
of protein density 1.35 g/cc (But see also \cite{protein density}),
and the reported amino acid frequencies in representative proteins
\cite{amino acid frequencies}, the representative atomic number densities
for the elements H, C, N, O, and S may be computed. The results are
shown in Table 1.

\begin{table}
\begin{tabular}{|c|c|c|c|c|c|}
\hline 
Element & H & C & N & O & S\tabularnewline
\hline
\hline 
Number Density {[}atoms/$\mathrm{nm}^{3}${]} & 62 & 6.4 & 1.8 & 28 & 0.067\tabularnewline
\hline
\end{tabular}

\caption{Number density of atoms in a typical biological specimen}

\end{table}


\begin{thebibliography}{73}
\bibitem{Comment 3} In contrast to Ref. \cite{ent-assisted EM},
here we use the symbol $\varphi$ instead of $\theta$ for the quantum
phase angle, in order not to be confused with the scattering angle
$\theta$.
\bibitem{Comment 4} The present author feels that appropriate data
processing, which is not dissimilar to computational tomography, would
help solving this problem. However, this has not been demonstrated
as of now.
\bibitem{Comment 5} The 'Laplacian filter' \cite{ent-assisted EM}
computes the difference of the following two data: The first data
is the result of applying a Gaussian filter with the 'standard deviation
' $0.3\mathrm{nm}$ to the unprocessed data, while the second is the
same except that the standard deviation is $1.5\mathrm{nm}$. It was
done on figures (b) and (d) to extract high resolution components
only, in order to make a comparison with figures 4 (a) and (c).
\end{thebibliography}
\end{document}